\documentclass[10pt,journal]{IEEEtran}
\ifCLASSINFOpdf
\else
\fi

\usepackage{amsmath, graphicx}
\usepackage{amssymb}
\usepackage[hidelinks]{hyperref}
\usepackage{mathabx}
\usepackage{bbm, xcolor}
\usepackage{booktabs}
\usepackage{rotating}
\usepackage{flushend}

\newcommand{\etal}{\textit{et al.}}

\begin{document}
%
\title{Learned Scalable Video Coding For Humans and Machines}
%
%
%

\author{Hadi~Hadizadeh and
        Ivan V.~Baji\'c
\thanks{This work was supported in part by NSERC grants RGPIN-2021-02485 and RGPAS-2021-00038. \\H. Hadizadeh and I. V. Baji\'c are with the School of Engineering Science, Simon Fraser University, Burnaby, BC, V5A 1S6, Canada, e-mail: \{hadi\_hadizadeh,ibajic\}@sfu.ca.}
}

%
%

\pagenumbering{gobble}

\markboth{EURASIP J. Image and Video Processing 2024:41 \hspace{6.7cm} \url{https://doi.org/10.1186/s13640-024-00657-w}}%
{EURASIP J. Image and Video Processing 2024:41}
%

\maketitle




\begin{abstract}
Video coding has traditionally been developed to support services such as video streaming, videoconferencing, digital TV, and so on. The main intent was to enable human viewing of the encoded content. However, with the advances in deep neural networks (DNNs), encoded video is increasingly being used for automatic video analytics performed by machines. In applications such as automatic traffic monitoring, analytics such as vehicle detection, tracking and counting, would run continuously, while human viewing could be required occasionally to review potential incidents. To support such applications, a new paradigm for video coding is needed that will facilitate efficient representation and compression of video for both machine and human use in a scalable manner. In this manuscript, we introduce an end-to-end learnable video codec that supports a machine vision task in its base layer, while its enhancement layer, together with the base layer, supports input reconstruction for human viewing. The proposed system is constructed based on the concept of conditional coding to achieve better compression gains. Comprehensive experimental evaluations conducted on four standard video datasets demonstrate that our framework outperforms both state-of-the-art learned and conventional video codecs in its base layer, while maintaining comparable performance on the human vision task in its enhancement layer.
Implementation of the proposed system is available at \url{https://github.com/hadipardis/svc}.
\end{abstract}

\begin{IEEEkeywords}
video compression, video analytics, scalable coding, deep learning, coding for machines
\end{IEEEkeywords}

%
\IEEEpeerreviewmaketitle

\section{Introduction}
%
%
%
%
\IEEEPARstart{V}{ideo} analytics is a crucial technology that finds application in a wide range of fields, such as visual surveillance, traffic monitoring, and autonomous navigation. In many applications, input video is compressed and transmitted to the cloud for subsequent analysis. Depending on the specific application, there are several approaches that can be adopted. When machine-based video analytics is the sole focus, precomputed features can be compressed and transmitted to the cloud instead of the complete video, thus reducing the required bandwidth~\cite{vcm1}. Various hand-crafted and neural-network-based (learned) approaches have been developed to serve this purpose~\cite{mpeg_cdvs, mpeg_cdva}. However, if human viewing is also required, the input video must also be encoded and transmitted, increasing the complexity of the system. 

Another approach involves compressing the input video and transmitting it to the cloud, where it will be decoded and used for visual analysis or human viewing. Various video codecs have been proposed over the last few decades for this purpose, including High Efficiency Video Coding (HEVC)~\cite{hevc} and Versatile Video Coding (VVC)~\cite{vvc}. However, if human viewing is needed only occasionally, this approach is inefficient, as will be shown later in the paper.

The swift progress in deep learning technologies has led to the emergence of several learned image and video codecs that rely on deep neural networks (DNNs) to compete with established conventional counterparts like JPEG, HEVC, and VVC. Notable methods include~\cite{balle2018, minnen, cheng2020, dcvc, canf, anfic, ladune, ssf, dcvc_tcm}. Nevertheless, a majority of DNN-based codecs have focused on image and video compression for human viewing. On the other hand, DNNs are widely used in visual analytics, but their primary emphasis is not on compression~\cite{video_analytics}. An important recent development in this space is JPEG AI~\cite{jpegai}, a DNN-based image coding standard whose aim is to support both human viewing and machine analysis within a single unified bitstream. An advantage of JPEG AI over earlier coding standards is that it offers the opportunity to bypass input image reconstruction and feed entropy-decoded latent representation into a DNN-based back-end for analysis, thereby facilitating potential savings in terms of complexity. However, the JPEG AI standard itself does not offer the means for the encoder to distinguish machine task-relevant from task-irrelevant information. Hence, when human viewing is needed only occasionally, this approach faces similar challenges in terms of bit-efficiency as earlier coding standards, because task-irrelevant information is encoded into the bitstream.

Recently, new video standardization efforts -- MPEG-VCM (Video Coding for Machines)~\cite{mpeg_vcm} and MPEG-FCM (Feature Coding for Machines) -- have been initiated with the aim of establishing a standardized framework for video coding that caters to the needs of both human and machine vision. Various approaches have also been proposed for scalable image coding for multiple tasks~\cite{sic1,sic2,hyomin_sic}, including machine vision and human viewing. In these methods, the base layer usually supports the machine vision task (e.g., object detection) while the enhancement layer supports input reconstruction for human viewing.

As discussed in~\cite{hyomin_sic}, a typical machine vision pipeline involves image encoding, decoding, and the analysis of the decoded image by a machine vision model. This process can be represented by a Markov chain~\cite{Cover_Thomas_2006} $X\rightarrow Y \rightarrow \widehat{X} \rightarrow V$, where $X$ denotes the input image, $Y$ represents the latent representation of $X$, $\widehat{X}$ is the reconstructed image, and $V$ is the output of the machine vision model. Applying the data processing inequality~\cite{Cover_Thomas_2006} to this chain, we can infer that $I(Y;\widehat{X}) \geq I(Y;V)$, where $I(\cdot \,;\, \cdot)$ represents the mutual information \cite{Cover_Thomas_2006}. This suggests that machine vision may require less information (fewer bits, if optimally encoded) than input reconstruction. This aligns with scalable human-machine coding approaches mentioned above, where both base and enhancement layers are utilized for input reconstruction, but only the base layer (i.e., fewer bits) is used for machine vision.

In this paper, we present an end-to-end learned scalable human-machine video codec. The proposed codec is designed to support object detection in its base layer, and input reconstruction by using both base and enhancement layer. The proposed system utilizes the back-end of a well-known DNN-based object detection network (YOLOv5)~\cite{yolo5} to execute the object detection task through its base layer. However, the architecture of the proposed system is versatile, and it can be integrated with other machine vision models. Since object detection is a fundamental machine vision task that serves as a building block for more complex video analytics -- such as video surveillance, traffic monitoring, object tracking, smart home, security systems, and so on -- it was a suitable choice that serves two purposes: 1) it provides information for a variety of downstream analytics applications, and 2) it allows us to demonstrate the compression benefits of removing task-irrelevant information from the base layer, thereby achieving significant gains over conventional approaches when only machine analysis is required. In terms of compression, the proposed system is constructed based on the notion of conditional coding~\cite{dcvc, canf, dcvc_tcm, lccm}, which is theoretically more efficient 
than the widely-employed residual coding, as discussed in Section~\ref{sec:motivation}.

We motivate our codec design by analyzing the efficiency of lossy conditional coding compared to lossy residual coding. In Section \ref{sec:motivation}, we present rate-distortion analysis of lossy conditional coding. To our knowledge, such analysis has not been presented in the video coding literature thus far. Our proposed scalable video coding system consists of two layers: a base layer and an enhancement layer. For the base layer, we leveraged our previous work presented in \cite{lccm}. For the enhancement layer, we incorporated some modules from the codec introduced in~\cite{dcvc_tcm} and added a novel feature extraction and fusion model, as well as a new entropy model, to enable efficient coding of the enhancement layer conditioned on the base layer.

After building a compression system on the principles of conditional coding, we performed extensive experiments to assess its performance. Our experiments on four standard video datasets demonstrate that the base layer of the proposed system outperforms state-of-the-art conventional and learned video codecs in coding for object detection, while the enhancement layer provides comparable performance for human viewing to those obtained by the existing codecs. In fact, our proposed codec serves as a practical demonstration of the effectiveness of conditional coding.
The proposed codec is well suited for applications where automated machine vision needs to run continuously while human viewing is needed occasionally, which we demonstrate through break-even analysis. 

This paper is organized as follows. We 
review some of the related prior work on learned video coding for humans and machines, and discuss the differences relative to our present work in Section~\ref{sec:prior_work}. After that, we provide a rate-distortion analysis of conditional coding in Section~\ref{sec:motivation}, which is used in our codec design. The proposed system is presented in Section~\ref{sec:proposed}. The experimental results are provided in Section \ref{sec:experiments}, and the conclusions are drawn in Section \ref{sec:conclusions}.

\section{Related Work}
\label{sec:prior_work}
In recent years, a number of learned video codecs have emerged in the literature. Lu \etal~\cite{dvc} introduced Deep Video Compression (DVC), the first end-to-end learned video codec utilizing temporal predictive coding. Agustsson \etal~\cite{ssf} proposed an end-to-end video coding model based on a learning-driven motion compensation framework, where a warped frame generated by a learned flow map serves as the predictor for coding the current video frame. Liu \etal~\cite{liu} employed feature-domain warping in a coarse-to-fine approach for video compression, while Hu \etal~\cite{hu} leveraged deformable convolutions for feature warping.

Most traditional video codecs rely on residual coding. However, Ladune \etal~\cite{theo2,ladune} argued that conditional coding relative to a predictor is more efficient than residual coding when using the same predictor. Building on this concept, Ho \etal~\cite{canf} introduced Conditional Augmented Normalizing Flows for Video Coding (CANF-VC), a high-performance learned video codec that utilizes conditional coding for both motion and inter-frame coding. Building on CANF-VC's conditional coding engines, we developed a codec called Learned Conditional Coding Modes for Video Coding (LCCM-VC)~\cite{lccm}, which incorporates additional learned coding modes to further enhance compression performance. In~\cite{dcvc_tcm}, a method called DCVC-TCM was proposed, utilizing a set of temporal multiscale contexts as predictors for conditional coding of video frames.

The methods mentioned above were specifically designed for video coding for human viewing. However, a few learned scalable video coding techniques have been developed that cater to both human vision and machine analytics. For example,~\cite{hyomin_svc} presents a scalable human-machine video codec based on the concept of latent-space scalability~\cite{hyomin_latent}. This codec was designed using a hybrid approach that leverages DNN-based compression for intra coding, and a combination of DNN and conventional techniques for inter-frame coding. In~\cite{face_videos}, a human-centric video compression system for human and machines was proposed, which can restore faces with high perceptual quality for human viewing at very low bitrates while maintaining high recognition accuracy for machine analytics. A human-machine friendly video compression scheme (HMFVC) was proposed in~\cite{hmfvc}. This scheme supports action recognition in its semantic (base) layer and input reconstruction for human viewing in its enhancement layer. A deep scalable video coding for human and machine vision, called DeepSVC, was proposed in~\cite{deep_svc}, which supports a semantic layer for object detection, a structure layer for low-bitrate input reconstruction, and a texture layer for higher quality input reconstruction.

\color{black}

The proposed system differs from~\cite{hyomin_svc} in several ways. First,~\cite{hyomin_svc} is a hybrid codec that incorporates both DNN-based and conventional handcrafted video coding tools, while the system presented in this paper is an end-to-end learned codec. Second,~\cite{hyomin_svc} used only intra-coding in the base layer, while our proposed system utilizes inter-frame coding for both the base and enhancement layers. Third,~\cite{hyomin_svc} employed conventional residual inter-frame coding in the enhancement layer, while our proposed codec is built upon conditional coding. Finally,~\cite{hyomin_svc} used YOLOv3~\cite{yolo3} for object detection in the base layer while our codec uses the more recent YOLOv5~\cite{yolo5}. The difference with respect to~\cite{hmfvc} is twofold. First, the machine task in~\cite{hmfvc} is action recognition while ours is  object detection. Second,~\cite{hmfvc} is built upon residual coding whereas our codec utilizes conditional coding. The proposed system also differs from~\cite{face_videos} in that~\cite{face_videos} is designed specifically for face videos (e.g., videoconferencing), whereas our codec is applicable across diverse application scenarios. Finally, although our codec and~\cite{deep_svc} both consider object detection as the machine task, there are several key differences. First,~\cite{deep_svc} utilizes TROIAlign~\cite{troi_align} for object detection, whereas our codec is based on the popular YOLOv5 pipeline. Second, there are architectural differences in both the coding engines and the layer structure. Finally, unlike~\cite{deep_svc}, which applies conditional coding solely for enhancement, we employ conditional coding in both base and enhancement layers.

\section{Rate-Distortion
Analysis of Conditional Coding}
\label{sec:motivation}
In this section we provide motivation for our compression approach by analyzing residual and conditional coding from the rate-distortion point of view. We will use capital letters ($X$) to denote random variables and lowercase letters ($x$) for their realizations. 

Conditional coding has been argued  in~\cite{dcvc, canf, ladune,  dcvc_tcm,lccm} to offer the potential for better compression performance compared to residual coding. However, these studies offered only arguments based on lossless compression. Specifically, if $X$ is the input to be compressed and $Y$ is its predictor, then it can be shown~\cite{lccm} that
\begin{equation}
    H(X|Y) \leq H(X-Y),
    \label{eq:lossless_conditional}
\end{equation}
where $H(\cdot)$ is the entropy and $H(\cdot \, | \, \cdot)$ is the conditional entropy~\cite{Cover_Thomas_2006}. If we were interested in lossless coding, this would be enough to motivate the choice of a conditional codec over a residual one, since entropy is the limit of lossless compression. However, most practical codecs are lossy, so we need to employ rate-distortion theory to argue in favor of conditional versus residual coding. In this section, we offer such analysis, which can be considered as the extension of~(\ref{eq:lossless_conditional}) into the realm of rate-distortion.

\begin{figure}[t]
    \centering
    \includegraphics[width=\columnwidth]{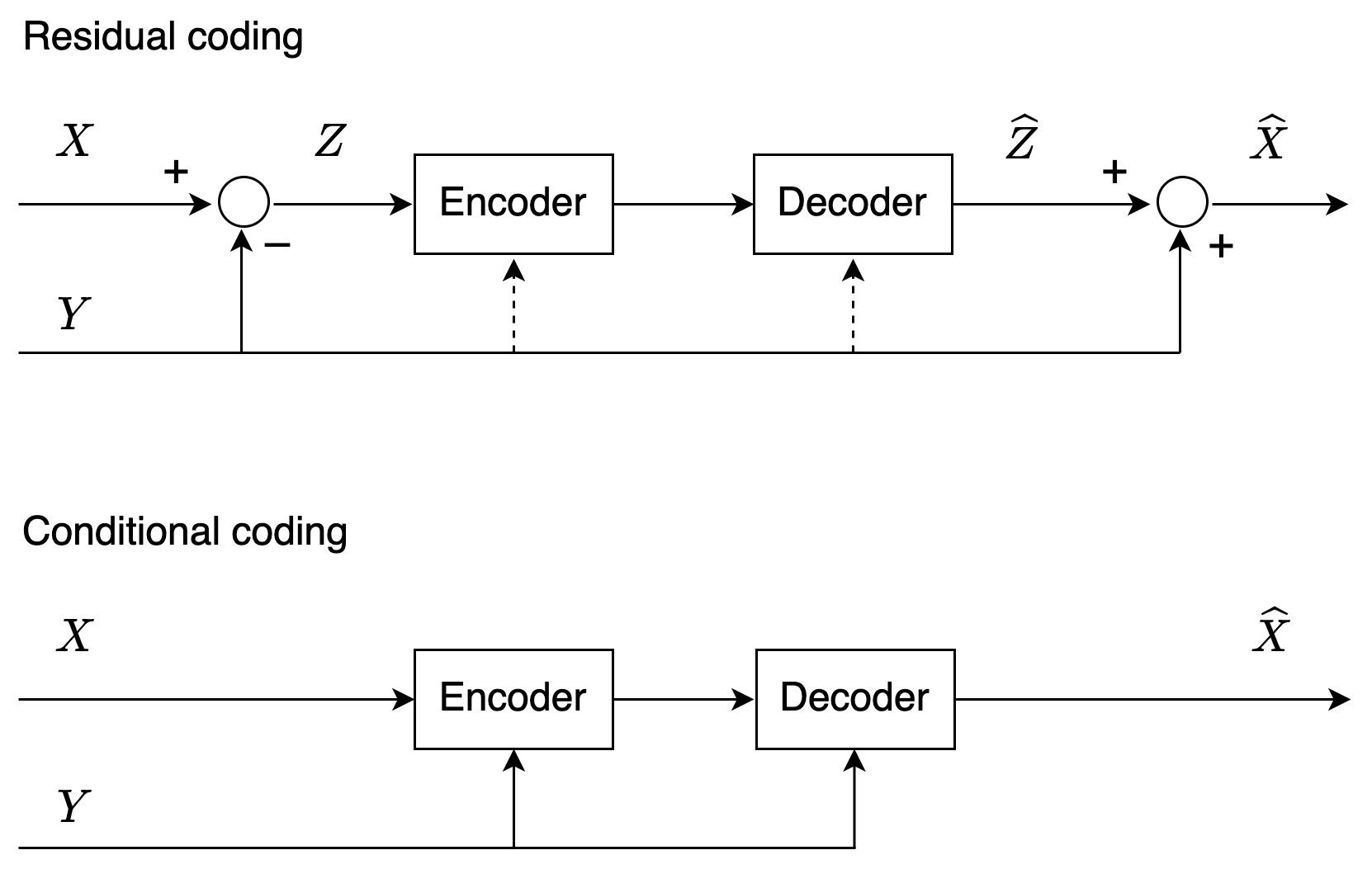}
    \caption{Top: residual coding. Dashed lines are not used in residual coding, they are shown to aid the analysis in the text. Bottom: conditional coding.}
    \label{fig:residual_conditional}
\end{figure}

First, we need to recall a few concepts from rate-distortion theory~\cite{Cover_Thomas_2006}. Let $d(x,\hat{x})$ be a distortion measure between an input $x$ and its reconstruction $\hat{x}$. The expected distortion between input $X$ and its reconstruction $\widehat{X}$ is computed as
\begin{equation}
    \mathbb{E}\left[d(X,\widehat{X})\right] = \sum_{x,\hat{x}} p_X(x)\cdot p_{\widehat{X}|X}(\hat{x}|x)\cdot d(x,\hat{x}),
    \label{eq:expected_distortion}
\end{equation}
where $p_X(x)$ is the input distribution and $p_{\widehat{X}|X}(\hat{x}|x)$ is the conditional distribution of the reconstruction given the input, also known as the \emph{quantizer}. The input distribution is not under our control, but the quantizer is. For a given distortion level $D>0$, the set of feasible quantizers for $X$ that achieve expected distortion of at most $D$ is denoted 
\begin{equation}
    \mathcal{P}_X(D)=\left\{p_{\widehat{X}|X}(\hat{x}|x) \, : \, \mathbb{E}\left[d(X,\widehat{X})\right] \leq D \right\}.
\end{equation}
The rate distortion function $R(D)$ is the minimum achievable rate for encoding input $X$ while incurring expected distortion of at most $D$. It is given by:
\begin{equation}
    R(D) = \min_{p_{\widehat{X}|X}(\hat{x}|x) \in \mathcal{P}_X(D)} I(X;\widehat{X}),
    \label{eq:RD_function}
\end{equation}
where $I(\cdot;\cdot)$ is the mutual information~\cite{Cover_Thomas_2006}. Note that in rate-distortion theory, rate is governed by mutual information rather than entropy, which is why~(\ref{eq:lossless_conditional}) cannot be used directly to motivate conditional lossy compression.

\begin{figure*}
\centering
\includegraphics[scale=0.47]{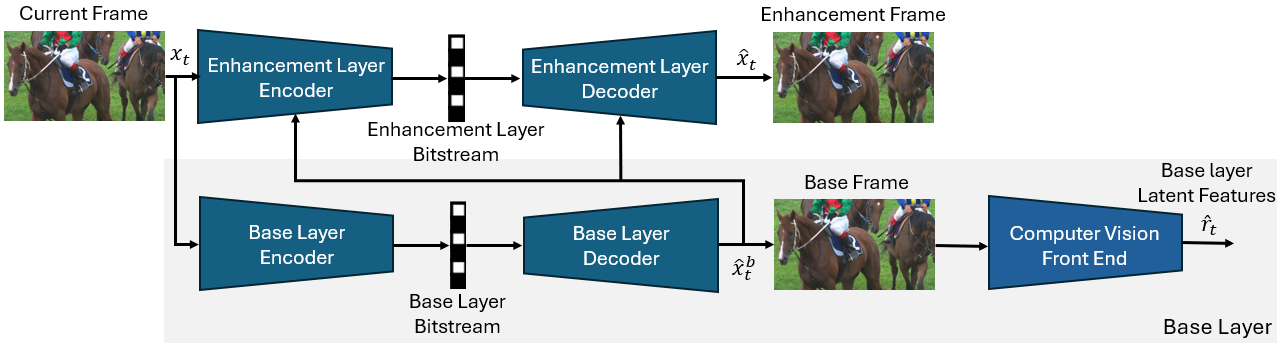}
\caption{The block diagram of the proposed learned scalable video compression system. The input video frame is $x_t$. The base layer has two outputs: the  base frame $\hat{x}_t^b$ and the latent features $\hat{r}_t$ by which the computer vision task is performed. The enhancement layer has only one output, enhancement frame, which is the reconstructed frame $\hat{x}_t$.}
\label{fig:flowchart}
\end{figure*}

Consider the two coding schemes illustrated in Fig.~\ref{fig:residual_conditional}, where the top part shows residual coding and the bottom part shows conditional coding. In residual coding, prediction $Y$ is subtracted from the input $X$ to create $Z=X-Y$, which is then encoded and reconstructed as $\widehat{Z}$. Finally, input reconstruction is obtained as $\widehat{X}=\widehat{Z}+Y$. If the distortion is shift-invariant\footnote{Many conventional distortion measures, including norm-based measures, are shift-invariant.} then $d(x,\hat{x}) = d(x-y,\hat{x}-y) = d(z,\hat{z})$, so $\mathbb{E}[d(X,\widehat{X})] = \mathbb{E}[d(Z,\widehat{Z})]$. Hence, the rate-distortion function for residual coding of $X$ is 
\begin{equation}
    R_r(D) = \min_{p_{\widehat{Z}|Z}(\hat{z}|z) \in \mathcal{P}_Z(D)} I(Z;\widehat{Z}),
    \label{eq:RD_residual}
\end{equation}
where subscript $r$ indicates residual coding and
\begin{equation}
    \mathcal{P}_Z(D)=\left\{p_{\widehat{Z}|Z}(\hat{z}|z) \, : \, \mathbb{E}\left[d(X,\widehat{X})\right] \leq D \right\}.
\end{equation}
Meanwhile, by applying (\ref{eq:RD_function}) to the input $(X,Y)$ and considering the case where the distortion is $0$ everywhere on the support of $Y$, we arrive at the rate-distortion function for conditional coding of $X$ relative to $Y$ (Fig.~\ref{fig:residual_conditional} bottom):
\begin{equation}
    R_c(D) = \min_{p_{\widehat{X}|X,Y}(\hat{x}|x,y) \in \mathcal{P}_{X|Y}(D)} I(X;\widehat{X}|Y),
    \label{eq:RD_conditional}
\end{equation}
where subscript $c$ indicates conditional coding,  $I(\cdot;\cdot|\cdot)$ is the conditional mutual information~\cite{Cover_Thomas_2006}, and 
\begin{equation}
    \mathcal{P}_{X|Y}(D)=\left\{p_{\widehat{X}|X,Y}(\hat{x}|x,y) \, : \, \mathbb{E}\left[d(X,\widehat{X})\right] \leq D \right\}.
\end{equation}
We expand $I(X;\widehat{X}|Y)$ as follows:
\begin{equation}
\begin{split}
    I(X;\widehat{X}|Y) &= H(X|Y) - H(X|\widehat{X},Y)\\
    &\stackrel{\text{(a)}}{=} H(X-Y|Y) - H(X-Y|\widehat{X},Y)\\
    &= I(X-Y;\widehat{X}|Y)\\
    &= H(\widehat{X}|Y) - H(\widehat{X}|Y,X-Y)\\
    &\stackrel{\text{(b)}}{=} H(\widehat{X}-Y|Y) - H(\widehat{X}-Y|Y,X-Y)\\
    &= I(X-Y;\widehat{X}-Y|Y)\\
    &\stackrel{\text{(c)}}{=}I(Z;\widehat{Z}|Y),
    \label{eq:MI_equivalence}
\end{split}
\end{equation}
where (a) (resp. (b)) follows from the fact that given $Y$, the only uncertainty in $X-Y$ (resp. $\widehat{X}-Y$) is due to $X$ (resp. $\widehat{X}$), and (c) follows from the fact that  $Z=X-Y$ and $\widehat{Z}=\widehat{X}-Y$. Moreover, due to these relationships, it should be clear that conditioned on $Y$, minimization over $p_{\widehat{X}|X,Y}(\hat{x}|x,y)$ is equivalent to minimization over $p_{\widehat{Z}|Z,Y}(\hat{z}|z,y)$. Combining this with~(\ref{eq:MI_equivalence}), we have that $R_c(D)$ from~(\ref{eq:RD_conditional}) can also be written as
\begin{equation}
     R_c(D) = \min_{p_{\widehat{Z}|Z,Y}(\hat{z}|z,y) \in \mathcal{P}_{Z|Y}(D)} I(Z;\widehat{Z}|Y),
    \label{eq:RD_conditional_Z}
\end{equation}
where 
\begin{equation}
    \mathcal{P}_{Z|Y}(D)=\left\{p_{\widehat{Z}|Z,Y}(\hat{z}|z,y) \, : \, \mathbb{E}\left[d(X,\widehat{X})\right] \leq D \right\}.
\end{equation}
Hence, conditional coding of $X$ relative to $Y$ achieves the same rate distortion as conditional coding of $Z$ relative to $Y$. This is shown in Fig.~\ref{fig:residual_conditional}(top), where the dashed line indicates conditional coding, mimicking Fig.~\ref{fig:residual_conditional}(bottom). 

Next, we turn to the seminal paper on conditional rate-distortion~\cite{ConditionalRD_IT1973}. In the notation used in~\cite{ConditionalRD_IT1973}, we have that our $R_r(D) = R_Z(D)$ and our $R_c(D) = R_{Z|Y}(D)$. Theorem 3.1 in~\cite{ConditionalRD_IT1973} shows that $R_{Z|Y}(D) \leq R_Z(D)$, with equality if and only if $Y$ and $Z$ are independent. Hence, we also have that
\begin{equation}
    R_c(D) \leq R_r(D),
    \label{eq:RD_inequality}
\end{equation}
meaning that for any given distortion level $D>0$, optimal conditional coding achieves the rate no higher than optimal residual coding. According to~\cite{ConditionalRD_IT1973}, the equality in~(\ref{eq:RD_inequality}) would hold if and only if $Y$ and $Z=X-Y$ were independent, i.e., if $Y$ and $X$ were independent. Hence, so long as the side information $Y$ provides at least some 
information about $X$,~(\ref{eq:RD_inequality}) becomes a strict inequality. In practice, however, we are generally dealing with sub-optimal codecs, so~(\ref{eq:RD_inequality}) might not always hold for practical codecs. 

Equation~(\ref{eq:RD_inequality}) is the rate-distortion analog of~(\ref{eq:lossless_conditional}). In fact,~(\ref{eq:lossless_conditional}) provides the result for the special case $D=0$ (lossless coding), while~(\ref{eq:RD_inequality}) extends this to $D>0$.  Intuition behind it is shown in  Fig.~\ref{fig:residual_conditional}(top), where the dashed lines imply that the conditional codec for $Z$ can, at the very least, ignore $Y$, and thereby match any performance that residual codec can achieve.

\section{Proposed Method}
\label{sec:proposed}
In this section, we present our proposed scalable video coding system, which is composed of a base layer and an enhancement layer. We first present an overview of the proposed codec in Section~\ref{sec:problem}. We then describe the structure of the base layer in Section~\ref{seq:base}, followed by the structure of the enhancement layer in Section~\ref{seq:enh}. The overall block diagram of the proposed system is depicted in Fig.~\ref{fig:flowchart}.

\begin{figure*}
\centering
\includegraphics[scale=0.6]{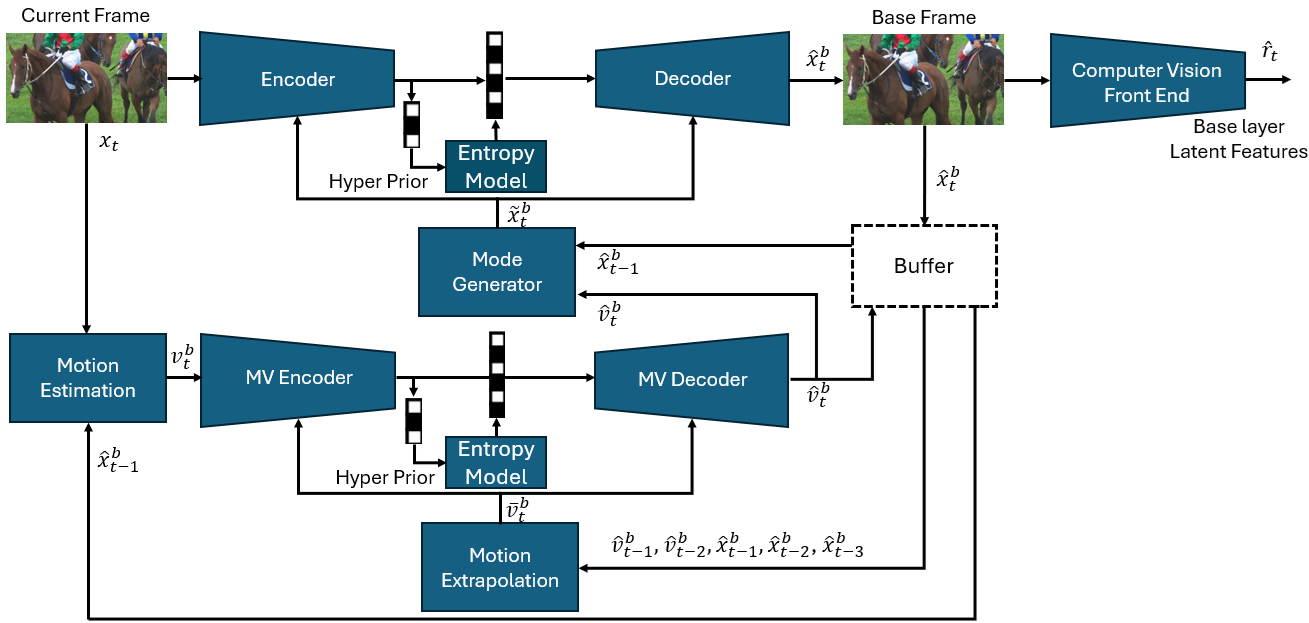}
\caption{The block diagram of the base layer of the proposed system. The base layer has two outputs: the reconstructed base frame $\hat{x}_t^b$ and the latent features $\hat{r}_t$ by which the computer vision task is performed. Our base layer's architecture is the same as our previous work LCCM-VC~\cite{lccm}. The Mode Generator is explained in more details in Fig.~\ref{fig:mode_generator}.}
\label{fig:flowchart_base_layer}
\end{figure*}

\subsection{Overview}
\label{sec:problem}
Let $x_{1:T} \in \mathbb{R}^{T\times 3\times H\times W}$ denote a raw (RGB) video sequence of $T$ frames $\{x_t,~ t=1,\cdots, T\}$ each of width $W$ and height $H$. The aim is to design an end-to-end learnable system to code the input video in a scalable manner that caters to both machine vision and human viewing. We design a system whose base layer facilitates object detection, intended for machine usage, while the enhancement layer, together with the base layer, facilitates reconstruction of the input video for human viewing. Although our scalable system is designed to support object detection, it can be re-trained to support any other machine vision task in the base layer. We chose object detection because it forms the basis for more complex machine vision scenarios such as video surveillance, tracking, traffic monitoring, etc. In our design, we exploit the concept of conditional coding which, as shown in the previous section, holds a theoretical advantage over the more common residual coding.
In summary, the proposed system operates as follows:
\begin{itemize}
    \item The input video frame is initially coded by the base layer to generate a reconstructed (decoded) base frame.
    \item The reconstructed base frame is then utilized by a computer vision model to perform the designated machine vision task (e.g., object detection).
    \item Simultaneously, the reconstructed base frame serves as a predictor in the enhancement layer for conditional coding of the input video frame, resulting in a reconstructed frame optimized for human viewing.
\end{itemize}

Specifically, the base layer encoder codes every video frame, $x_t$, into a compressed base layer bitstream. The base layer decoder then decodes the compressed bitstream to obtain a decoded RGB frame, $\hat{x}_t^b$, also referred to as the decoded \textit{base frame}. This frame is utilized by a computer vision (CV) model (in our case YOLOv5~\cite{yolo5}) for video analysis (in our case object detection). Furthermore, $\hat{x}_t^b$ also helps the enhancement layer perform the reconstruction of the input frame for human viewing. The base layer is trained to maximize object detection accuracy while minimizing the rate of the base layer's bitstream. 
Note that the output of the base layer is an RGB frame. We designed it this way to ensure seamless integration with any existing computer vision network in a plug-and-play manner, eliminating the need for architectural modifications. For instance, this design enables us to easily replace YOLO~\cite{yolo3} with any object detector, such as SSD~\cite{ssd}, without changing their input architecture. These detectors expect an RGB image in their input, and our base layer outputs an RGB image. 

The input video frame, $x_t$, is also fed into the enhancement layer alongside the base frame, $\hat{x}_t^b$. The enhancement layer's encoder then compresses the input frame conditionally to generate a compressed bitstream, which the enhancement layer's decoder utilizes to reconstruct an output frame, $\hat{x}_t$. The enhancement layer is trained end-to-end to minimize both the distortion between $x_t$ and $\hat{x}_t$, as well as the rate of the enhancement layer's bitstream. In the proposed system, the base layer is trained first, followed by training the enhancement layer while keeping the base layer frozen. In scalable human-machine image coding, this strategy has recently been shown to be advantageous over parallel training of the base and enhancement layer~\cite{foroutan2023base}, so we use it for our proposed video codec as well. 

It is important to note that in practice, if the sole objective is to perform the computer vision task, there is no need to reconstruct the input frame. That is, the enhancement layer's bitstream does not need to be created because $\hat{x}_t$ does not need to be reconstructed at the decoder. If human viewing is required, the enhancement layer can be employed and $\hat{x}_t$ reconstructed. Many video analytics applications, such as traffic monitoring or video surveillance, require continuous use of the analytics engine (base layer in our system), and only occasional human viewing (enhancement layer), for cases like emergencies or incident review.

\subsection{The Base Layer}
\label{seq:base}
As shown in Fig. \ref{fig:flowchart_base_layer}, the base layer consists of several components, which are described below. The base layer codec is built upon our previous work LCCM-VC~\cite{lccm}; the main difference is that the codec in~\cite{lccm} was trained to preserve the fidelity of the input frame under a rate constraint, while our base layer codec here is trained to preserve task-relevant features $\hat{r}_t$ under a rate constraint. 

\subsubsection{Motion Estimation, Extrapolation, and Coding} Initially, the input video frame, $x_t$, and the previous base frame, $\hat{x}_{t-1}^b$, are fed into a DNN-based motion estimation network. This network is designed to estimate the optical flow, denoted by $v_t^b$, between $x_t$ and $\hat{x}_{t-1}^b$. For this purpose, we use PWC-Net \cite{pwc}. The flow $v_t^b$ is then encoded by a DNN-based encoder, $V_{\pi}$, to obtain a compressed motion bitstream in the base layer. This stream is decoded by a DNN-based decoder, $V_{\pi}^{-1}$, to obtain a decoded flow, $\hat{v}_t^b$. For this purpose, we adopt the conditional coder from CANF-VC~\cite{canf}, which codes a signal conditioned on its predictor to achieve better rate-distortion performance. To obtain a predictor for conditional coding of $v_t^b$, we utilized the motion extrapolation network proposed in \cite{canf}. This network leverages the two previously-decoded flow maps, $\hat{v}_{t-2}^b$ and $\hat{v}_{t-1}^b$, as well as the three previously-decoded base frames, $\hat{x}_{t-3}^b$, $\hat{x}_{t-2}^b$, and $\hat{x}_{t-1}^b$, to generate a predictor, $\bar{v}_t^b$, for $v_t^b$.

The optical flow map $\hat{v}_{t}^b$ is used to warp the previously-decoded base frame $\hat{x}_{t-1}^b$ through a motion compensation network to obtain a predictor frame, $\bar{x}_t^b$, for conditional coding of the current frame $x_t$. Also, $\hat{v}_{t}^b$ is stored in a flow buffer to be used for motion extrapolation of the subsequent frames. We used the same motion compensation network as proposed in \cite{canf}.

\subsubsection{Mode Generator} 
\begin{figure}
\centering
\includegraphics[width=\columnwidth]{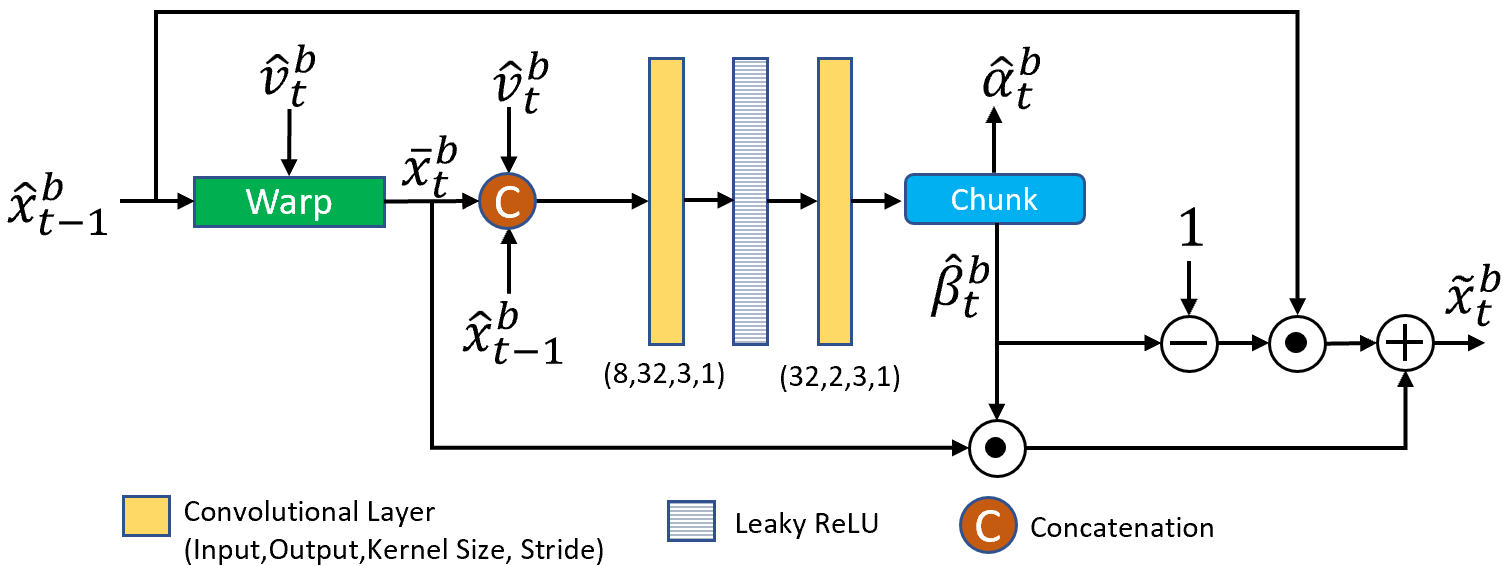}
\caption{The structure of the mode generator.}
\label{fig:mode_generator}
\end{figure}

To improve the coding performance of the base layer, we used the idea of learned conditional coding modes proposed in our earlier work on learned video coding, called LCCM-VC~\cite{lccm}. Specifically, the previously-reconstructed base frame $\hat{x}_{t-1}^b$, the motion-compensated frame $\bar{x}_t^b$, and the decoded flow $\hat{v}_t^b$ are fed to the mode generator from~\cite{lccm} to obtain two weight maps, $\hat{\alpha}_t^b$ and $\hat{\beta}_t^b$. Since these maps are produced from previously (de)coded data, they can be regenerated at the decoder without any additional bits. As shown in Fig. \ref{fig:mode_generator}, using $\hat{\beta}_t^b$, an enhanced predictor, $\tilde{x}_t^b$, is generated as follows:
\begin{equation}
    \label{eq:enhanced_predictor}
    \tilde{x}_t^b = \hat{\beta}_t^b \odot \bar{x}_t^b + (\mathbbm{1}-\hat{\beta}_t^b) \odot \hat{x}_{t-1}^b,
\end{equation}
where $\odot$ denotes Hadamard (element-wise) product and $\mathbbm{1}$ is the all-ones matrix. At each pixel location, the value of  $\tilde{x}_t^b$ is a weighted combination of $\bar{x}_t^b$ and $\hat{x}_{t-1}^b$, where the weight is defined by the corresponding value in $\hat{\beta}_t^b$. Hence, $\tilde{x}_t^b$ is a predictor for the current frame being coded, whose values are adaptively obtained from the previously-reconstructed base frame ($\hat{x}_{t-1}^b$) and the motion-compensated base frame ($\bar{x}_t^b$). After that, $\tilde{x}_t^b$ is multiplied by $\hat{\alpha}_t^b$ in a pixel-wise manner, and the obtained result, i.e. $\hat{\alpha}_t^b \odot \tilde{x}_t^b$, is fed as a predictor to the inter-frame coder described next. Note that, using (\ref{eq:enhanced_predictor}), the enhanced predictor can be considered as a linear combination of $\bar{x}_t^b$ and $\hat{x}_{t-1}^b$. When $\hat{\beta}_t^b \to 0$, it becomes equal to $\hat{x}_{t-1}^b$, and when $\hat{\beta}_t^b \to \mathbbm{1}$, it becomes equal to $\bar{x}_t^b$.

\subsubsection{Base Inter-frame Coder} The predictor $\hat{\alpha}_t^b \odot \tilde{x}_t^b$ is used for conditional coding of the current frame $x_t$. 
For this purpose, we utilized LCCM-VC~\cite{lccm}, which is an extension of 
CANF-VC~\cite{canf}. 
Using this framework, the conditional encoder of the base inter-frame coder, $G_{\pi}$, encodes $\hat{\alpha}_t^b \odot x_t$ using $\hat{\alpha}_t \odot \tilde{x}_t^b$ to produce a compressed bitstream, which is used to decode an image $\widecheck{x}_t^b$ via $G_{\pi}^{-1}$. Finally, the reconstructed base frame, $\hat{x}_t^b$, is generated as follows:
\begin{equation}
    \label{eq:decoded_input}
    \hat{x}_t^b = \widecheck{x}_t^b + (\mathbbm{1}-\hat{\alpha}_t^b) \odot \tilde{x}_t^b.
\end{equation}
 Note that, for pixels where $\hat{\alpha}_t^b \to 0$, $\hat{x}_{t}$ becomes equal to $\tilde{x}_{t}$, so the inter-frame coder does not need to code anything. This resembles the SKIP mode in conventional video coders, and depending on the value of $\hat{\beta}_t^b$, the system can copy directly from $\hat{x}_{t-1}^b$, $\bar{x}_{t}^b$, or a mixture of these two, to obtain $\tilde{x}_{t}^b$. When $\hat{\alpha}_t^b \to \mathbbm{1}$, only the inter-frame coder is used to obtain $\hat{x}_{t}$. In the limiting case when $\hat{\alpha}_t^b \to \mathbbm{1}$ and $\hat{\beta}_t^b \to \mathbbm{1}$, LCCM-VC~\cite{lccm} becomes CANF-VC~\cite{canf}. Hence, as discussed in \cite{lccm}, LCCM-VC has more flexibility and a larger number of conditional coding modes than CANF-VC.
The architecture of $G_{\pi}$ and $G_{\pi}^{-1}$ is similar to that of $V_{\pi}$ and $V_{\pi}^{-1}$.

\subsubsection{Task-Relevant Feature Extraction}
\label{sec:task}

The obtained base frame, $\hat{x}_t^b$, is stored in the base frame buffer for subsequent frame coding and is also fed to the base task network. In our work, this is YOLOv5~\cite{yolo5} for object detection, but in principle any (differentiable) network could be used here for any computer vision task. To enable unsupervised training of the base layer, we truncate the task network and utilize a copy of the resulting front-end of the network within the base layer, as described next. Specifically, we feed $\hat{x}_t^b$ to the cloned front-end network in the following manner:
\begin{equation}
\label{eq:feature_tensor}
\hat{r}_t = F^{\textrm{front-end}}_{\textrm{trainable}}(\hat{x}_t^b),
\end{equation}
where $F^{\textrm{front-end}}_{\textrm{trainable}}(\cdot)$ denotes the cloned front-end network, which is trainable during the training process alongside the rest of the base layer, and $\hat{r}_t$ is the resultant feature tensor. Then $\hat{r}_t$ is fed to the back-end part of the original pre-trained task network,  $F^{\textrm{back-end}}_{\textrm{original}}(\hat{r}_t)$, where $F^{\textrm{back-end}}_{\textrm{original}}(\cdot)$ is not trained during the training process. 

The primary advantage of this methodology is to eliminate the need for ground-truth labels for the base layer. The amount of labeled data for video (especially uncompressed video, which is needed in our application) is very limited, and it makes sense to use it all for testing. So instead, we train the base layer to match the features $r_t$ of the pre-trained network at a particular layer, where
\begin{equation}
\label{eq:feature_tensor_actual}
r_t = F^{\textrm{front-end}}_{\textrm{original}}(x_t),
\end{equation}
and this way, we don't need to use labelled video in training. 

The selection of the layer for feature matching will depend on the application constraints. The rate-distortion results in~\cite{harell2022PCS,harell2023TPAMI} suggest that better performance is achieved when matching deeper layers. However, this requires making $F^{\textrm{front-end}}_{\textrm{trainable}}(\cdot)$ larger, which in turn increases the complexity of the base layer. For the experiments in this study, we chose the first five layers of YOLOv5 as the front-end to demonstrate the performance of the system. In practice, the optimal choice will depend on the computational resources, allowed latency, available bandwidth, etc., in a particular application scenario. 

\subsubsection{Base Layer Loss Function} The loss function for training the proposed base layer is defined as follows:
\begin{equation}
\label{eq:base_loss}
L_{base}(t) = R_{base}(t) + \lambda_{base}~\textrm{MSE}(r_t, \hat{r}_t)
\end{equation}
where MSE$(r_t, \hat{r}_t)$ is the mean squared error between features derived from an uncompressed frame $(r_t)$ and features decoded from the base layer $(\hat{r}_t)$, $\lambda_{base}$ is the Lagrange multiplier, and $R_{base}(t)$ is the total rate of the base layer for coding $x_t$, which is defined as:
\begin{equation}
    R_{base}(t) = R_{base}^{\textrm{motion}}(t) + R_{base}^{\textrm{signal}}(t),
\end{equation}
where $R_{base}^{\textrm{motion}}(t)$ is the rate of the motion bitstream produced by coding $v_t^b$ conditioned on $\bar{v}_t^b$ plus the rate associated for its hyper prior, and $R_{base}^{\textrm{signal}}(t)$ is the rate for  coding $x_t$ conditioned on $\bar{x}_t^b$ plus the rate associated for its hyper prior. As seen from the above equations, the goal of the base-layer codec is to preserve task-relevant features while minimizing the bitrate. This means that the base layer codec is not asked to reconstruct (or code) task-irrelevant information, which usually results in a lower bitrate compared to conventional coding approaches that try to preserve the overall input fidelity.

\begin{figure}
\centering
\includegraphics[width=\columnwidth]{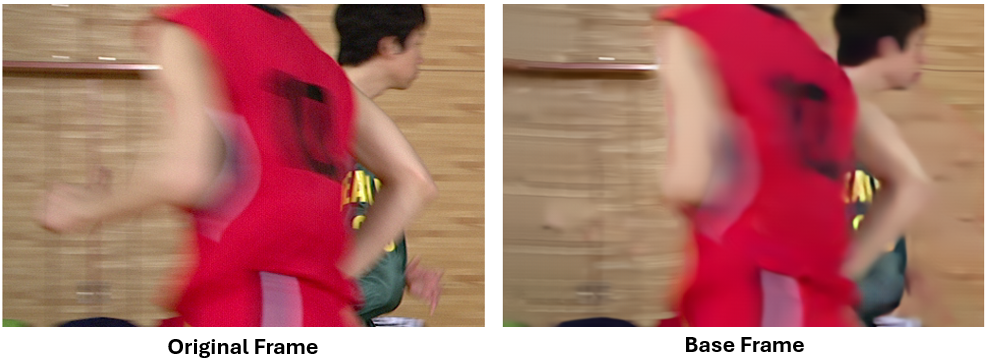}
\caption{A visual example of a portion of the base frame at 0.11 bpp. The original frame $x_t$ (left) and the base frame $\hat{x}_t^b$ (right).}
\label{fig:base_image}
\end{figure}

A visual example of a portion of the base frame ($\hat{x}_t^b$) in the trained system is shown in Fig. \ref{fig:base_image}, along with the original frame ($x_t$). As is evident from this figure, the base frame omits certain details of the original frame, which are deemed unnecessary for the object detection task, in order to curtail the rate. Consequently, the base layer can be encoded at a significantly lower rate than the original frame.

\subsection{The Enhancement Layer}
\label{seq:enh}
\begin{figure*}
\centering
\includegraphics[scale=0.5]{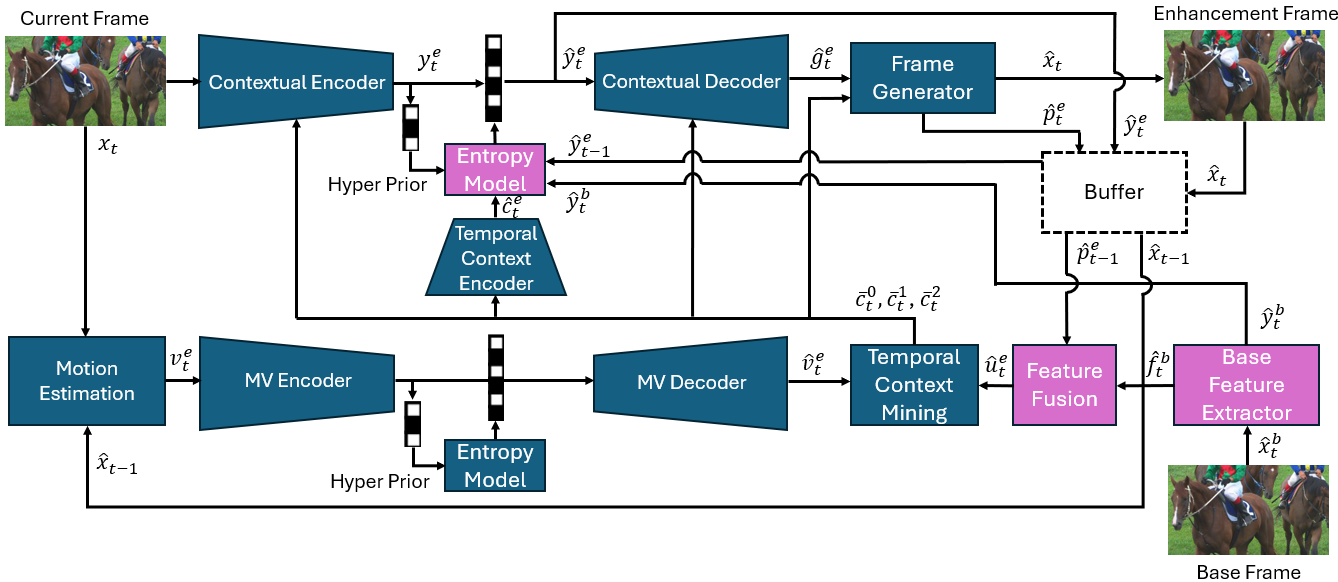}
\caption{The block diagram of the enhancement layer of the proposed system. It gets the current frame $x_t$ and the base frame $\hat{x}_t^b$ as inputs and generates the enhancement frame $\hat{x}_t$ as output. Our proposed enhancement layer is built upon DCVC-TCM \cite{dcvc_tcm}. The purple blocks are our proposed components.}
\label{fig:flowchart_enh_layer}
\end{figure*}

To implement the enhancement layer of the proposed system, we used the conditional coding-based framework proposed in~\cite{dcvc_tcm} as the basis. However, we made some modifications to fit it into our scalable codec, as will be discussed in the following subsections. The components of the enhancement layer are shown in Fig.~\ref{fig:flowchart_enh_layer}. In summary, the enhancement layer works as follows:
\begin{itemize}
    \item The current frame and the previously decoded video frame from the enhancement layer are used to estimate a motion flow map, which is subsequently encoded by a motion coder.
    \item The base frame generated by the base layer is processed by a base feature extractor, and the resulting base features are fused with the latent representations of the previously reconstructed frame in the enhancement layer.
    \item The fused features from the previous step, combined with the decoded motion flow map, are input into a temporal context mining module to learn motion-aligned temporal contexts. These contexts capture a multi-scale representation of the current frame, each serving as a predictor for the frame within the latent space.
    \item The current frame is then encoded by a conditional contextual codec, which uses the generated temporal contexts as predictors. This codec employs an entropy model with parameters estimated from the temporal contexts, along with the latent representations of the base frame and the previously decoded frame in the enhancement layer. 
    \item The contextual decoder, together with a frame generator, reconstructs the encoded video frame in the enhancement layer,  which is called ``Enhancement Frame.''
\end{itemize}
We will now provide a detailed description of each component within the enhancement layer.

\subsubsection{Motion Estimator and Coder} The current frame $x_t$ and the previously-reconstructed frame in the enhancement layer, $\hat{x}_{t-1}$, are fed to a motion estimation network to estimate a flow map, $v_t^e$, in the enhancement layer. Similar to the base layer, we also use PWC-Net~\cite{pwc} for motion estimation in the enhancement layer. The estimated flow map is then coded by a hyper-prior-based encoder~\cite{balle2018}, $P_{\pi}$, to obtain the compressed motion bitstream in the enhancement layer. During training, the hyper-prior-based encoder provides the rate estimate for the motion information in the enhancement layer, $R_{enh}^{\textrm{motion}}(t)$. The motion bitstream is then decoded by the corresponding hyper-prior-based decoder, $P_{\pi}^{-1}$, to obtain the decoded flow map, $\hat{v}_t^e$, which provides an estimated motion vector for each pixel within the current frame in the enhancement layer. 

\subsubsection{Base Feature Extractor and Fusion}
\begin{figure}
\centering
\includegraphics[scale=0.45]{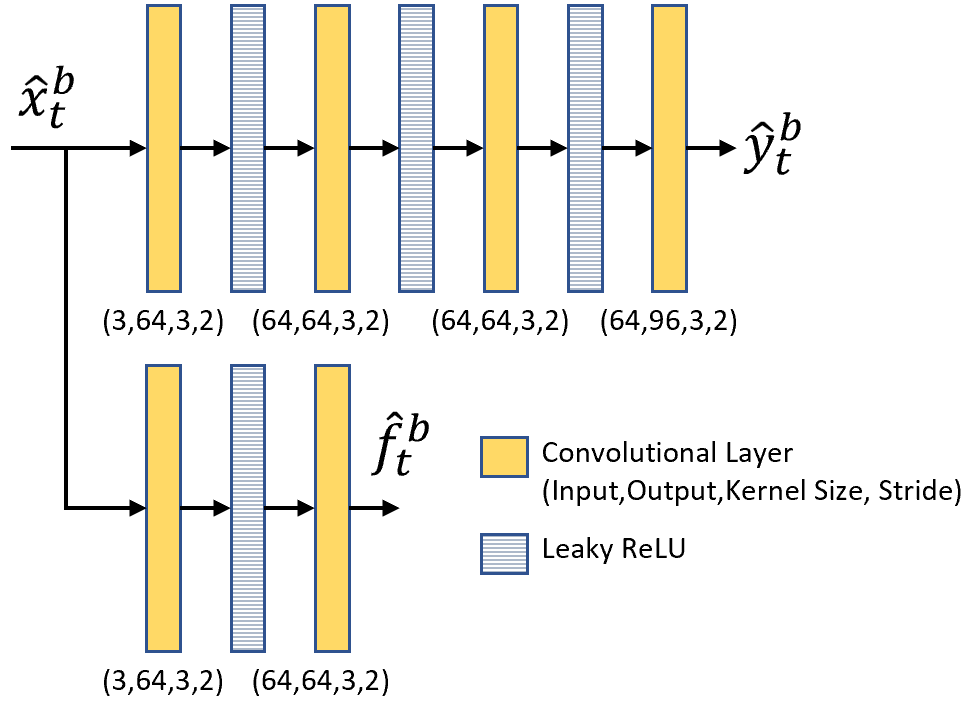}
\caption{The structure of the base feature extractor network.}
\label{fig:base_feature_extractor}
\end{figure}

\begin{figure}
\centering
\includegraphics[scale=0.45]{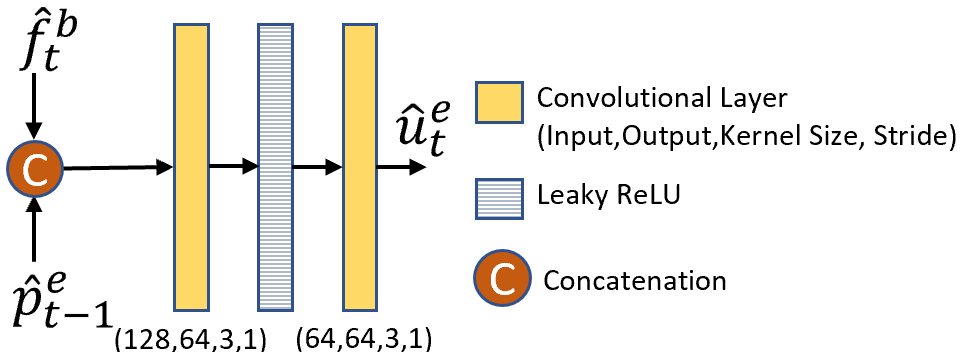}
\caption{The structure of the feature fusion network used in the enhancement layer.}
\label{fig:fusion_base}
\end{figure}

As shown in Fig. \ref{fig:base_feature_extractor}, the base frame $\hat{x}_t^b$ is first fed to a base feature extractor network  to obtain two output feature tensors: $\hat{f}_t^b$ and $\hat{y}_t^b$. Note that the base feature extractor network is different from the front-end of the task model. The first feature tensor, $\hat{f}_t^b$, is then fused with the features of the previous frame reconstructed by the frame generator, $\hat{p}_{t-1}^e$, to generate a new feature tensor $\hat{u}_t^e$. The feature fusion is performed by a simple network with two convolutional layers whose structure is shown in Fig. \ref{fig:fusion_base}. The second output of the base feature extractor, $\hat{y}_t^b$, is used by the entropy model, which is described in Section \ref{sec:entropy_model}.

\subsubsection{Temporal Context Mining} \label{sec:TCM} 
\begin{figure}
\centering
\includegraphics[scale=0.45]{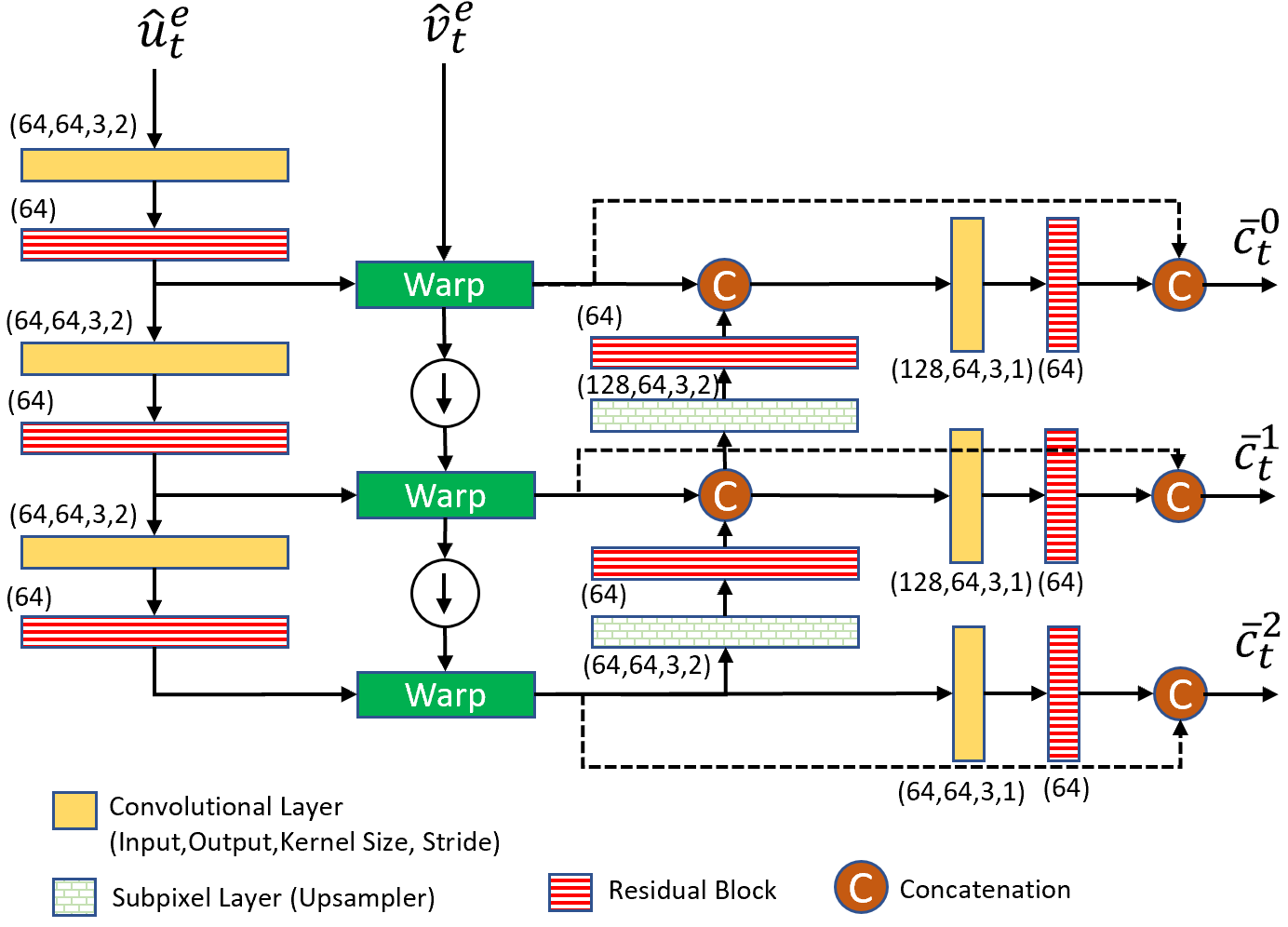}
\caption{The structure of the temporal context mining (TCM) network.}
\label{fig:tcm}
\end{figure}

\begin{figure}
\centering
\includegraphics[scale=0.45]{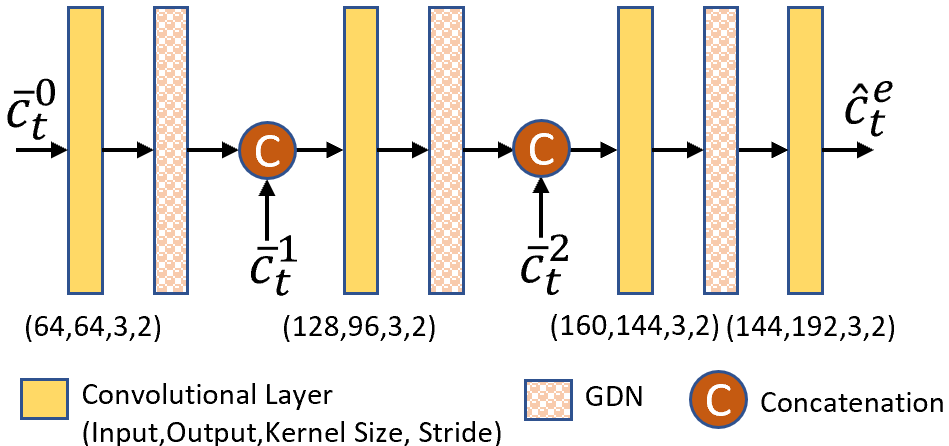}
\caption{The structure of the temporal prior network.}
\label{fig:temporal_prior}
\end{figure}

The goal of the Temporal Context Mining (TCM) module is to generate a set of multi-scale temporal contexts as predictors for conditional coding of the current frame using the contextual encoder. The structure of TCM is shown in Fig.~\ref{fig:tcm}. To generate the temporal contexts to encode the current frame, $\hat{u}_t^e$ and the decoded flow map $\hat{v}_t^e$ are fed to the TCM module proposed in~\cite{dcvc_tcm}. TCM uses $\hat{v}_t^e$ and warps $\hat{u}_t^e$ at three different scales to obtain three contexts $\{\bar{c}_t^0, \bar{c}_t^1, \bar{c}_t^2\}$, each at a different scale. These contexts provide rich latent temporal information for coding the current frame in the enhancement layer. After that, these three contexts are combined through a temporal prior network whose structure is shown in Fig.~\ref{fig:temporal_prior}. The output of this network is $\hat{c}_t^e$, which conveys the temporal context of the current frame. 

In~\cite{dcvc_tcm}, $\hat{p}_{t-1}^e$ is used directly as input to the TCM module. However, in our design, we fuse $\hat{p}_{t-1}^e$ with $\hat{f}_t^b$ through the feature fusion network, and the resulting output, $\hat{u}_t^e$, is used as input to the TCM module. In this way, the generated temporal contexts leverage the information provided in both the base frame and the previous reconstructed frame in the enhancement layer.

\subsubsection{Contextual Encoder} 
\begin{figure*}
\centering
\includegraphics[scale=0.45]{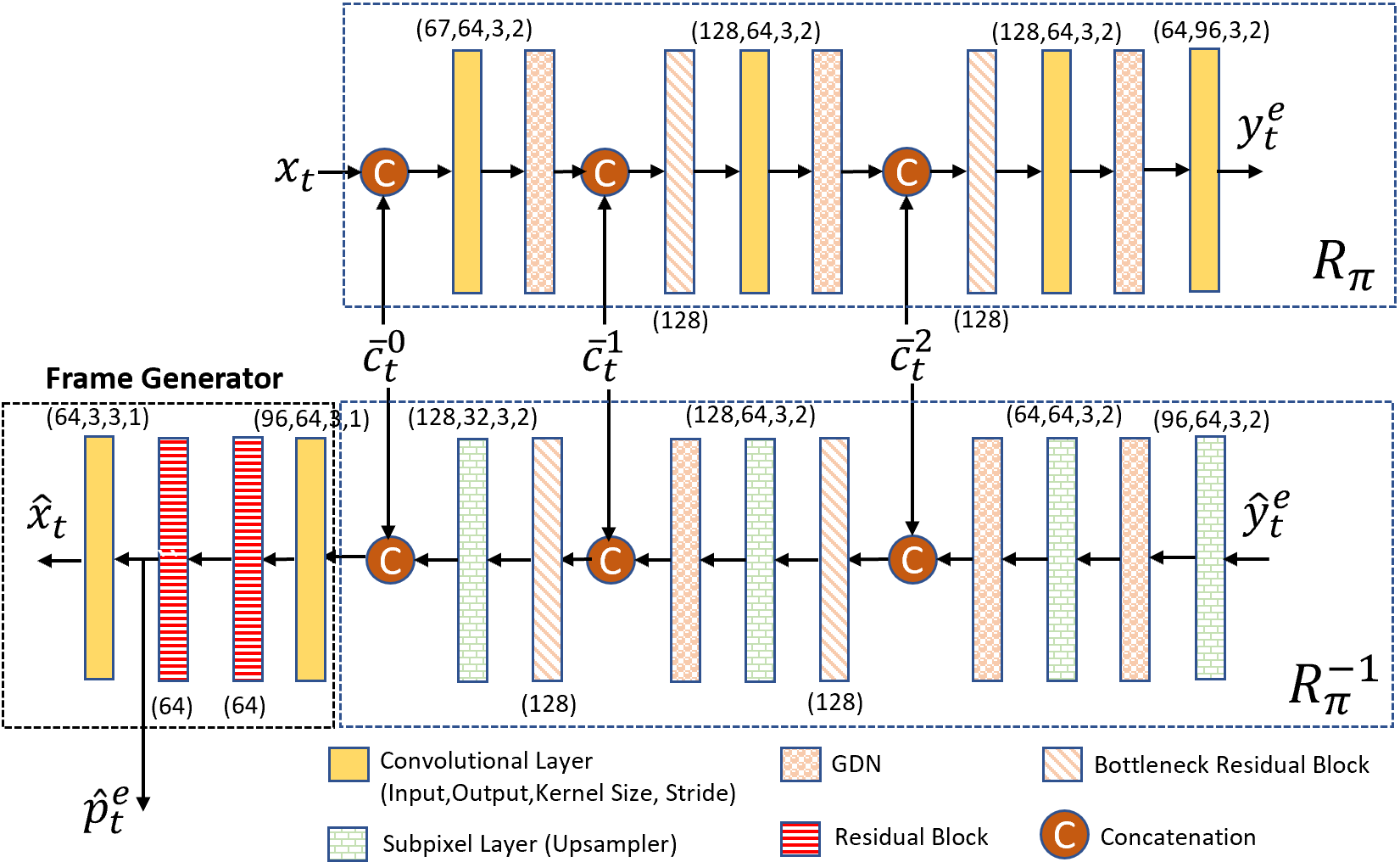}
\caption{The structure of the context encoder $R_{\pi}$, the context decoder $R_{\pi}^{-1}$, and the frame generator. GDN represents the Generalized Divisive Normalization layer \cite{GDN}.}
\label{fig:enh_layer}
\end{figure*}

The structure of the contextual encoder is shown in Fig. \ref{fig:enh_layer}. As shown in this figure, the current frame $x_t$ is encoded by a conditional contextual encoder, $R_{\pi}$, to produce a latent representation $y_t^e$. The conditional contextual encoder uses the three temporal contexts $\{\bar{c}_t^0, \bar{c}_t^1, \bar{c}_t^2\}$ for conditional coding of $x_t$. These contexts are produced by the TCM module described in Section~\ref{sec:TCM}. The obtained latent representation $y_t^e$ is then quantized and compressed by an arithmetic encoder whose symbol probabilities are estimated by a hyper prior-based entropy model \cite{balle2018}. Using this approach, a compressed contextual bitstream is produced along with a side compressed bitstream for signaling the hyper-priors $h_t^e$. The contextual encoder uses an entropy model, which is described in Section~\ref{sec:entropy_model} below.

\subsubsection{Contextual Decoder} The structure of the contextual decoder is shown in Fig.~\ref{fig:enh_layer}. The contextual bitstream produced by the contextual encoder is  decompressed at the decoder by an arithmetic decoder to obtain $\hat{y}_t^e$, which is then fed to a conditional contextual decoder, $R_{\pi}^{-1}$. Similarly, the hyper-priors of the current frame are decoded from the side bitstream to obtain the temporal hyper-prior  $\hat{h}_t^e$.
A copy of $\hat{y}_t^e$ is stored in the generalized decoded picture buffer (DPB) to be used for encoding the subsequent frames. $R_{\pi}^{-1}$ uses the three temporal contexts and transforms $\hat{y}_t^e$ into a new latent representation $\hat{g}_t^e$, which is then fed to the frame generator to reconstruct the transmitted frame, as will be described next. 

\subsubsection{Frame Generator} 
The frame generator passes its input tensor $\hat{g}_t^e$ through a number of convolutional layers and residual blocks as in~\cite{dcvc_tcm}, and reconstructs the input frame as $\hat{x}_t$. As in~\cite{dcvc_tcm}, the output of the layer before the last layer in the frame generator, $\hat{p}_t^e$, is also stored in the DPB along with $\hat{x}_t$. The feature tensor $\hat{p}_t^e$ is subsequently used by the TCM network to produce the temporal contexts.

\subsubsection{Entropy Model} \label{sec:entropy_model} Similar to \cite{dcvc_tcm} and \cite{dcvc}, we use the factorized entropy model for the hyper-prior coder, and the Laplace distribution to model the quantized latent representations in the enhancement layer, $\hat{y}_t^e$. However, to improve the performance of the entropy model, we fuse the hyper priors of the current frame, $\hat{h}_t^e$, with the temporal prior $\hat{c}_t^e$, decoded latents from the previous enhancement frame $\hat{y}_{t-1}^e$ and base features of the current frame $\hat{y}_t^b$ through an entropy parameter estimation network. In \cite{dcvc_tcm}, only the hyper-prior $\hat{h}_t^e$ and the temporal prior $\hat{c}_t^e$ are fed to the entropy model. However, in our design, in addition to $\hat{h}_t^e$ and $\hat{c}_t^e$, we also feed $\hat{y}_t^b$ and $\hat{y}_{t-1}^e$ to the entropy model to enrich its input information. Moreover, we add the attention layer so the system can learn where to take the relevant information for entropy modelling. The proposed entropy model is implemented using an entropy estimation network, which is described next. \\

\begin{figure*}
\centering
\includegraphics[scale=0.4]{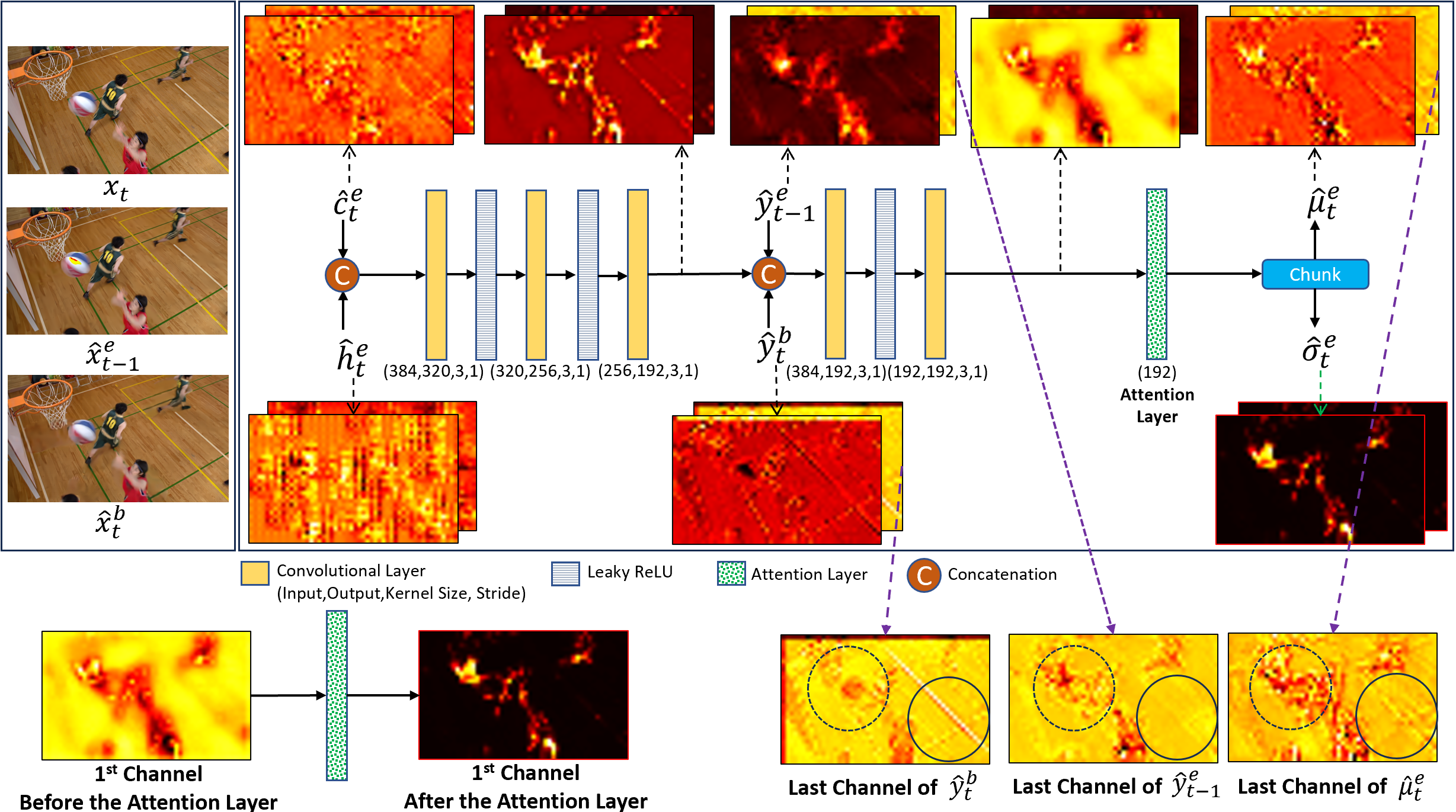}
\caption{The structure of the proposed entropy parameter estimation network. The latent feature maps are shown for a sample frame from BaskteballDrill (HEVC Class C).}
\label{fig:entropy_parameter}
\end{figure*}

\textbf{The Entropy Estimation Network:}
The structure of the entropy model is shown in Fig.~\ref{fig:entropy_parameter}. 
As shown in this figure, the hyper-prior $\hat{h}_t^e$ and the temporal prior $\hat{c}_t^e$ are first concatenated, and fed to three consecutive convolutional layers. The resulting feature tensor is then concatenated with $\hat{y}_{t-1}^e$ and $\hat{y}_t^b$, and the generated tensor is fed to two consecutive convolutional layers followed by an attention layer. The output of the attention layer is then chunked in half to produce two tensors, $\hat{\mu}_t^e$ and $\hat{\sigma}_t^e$, which are used to estimate the mean and variance of the Laplace distribution, respectively.

Sample frames from the BasketballDrill sequence (HEVC Class C) are also shown in Fig.~\ref{fig:entropy_parameter}: $x_t$ is the current (to-be-coded) frame at time $t$, $\hat{x}^e_{t-1}$ is the previously decoded frame in the enhancement layer, and $\hat{x}^b_t$ is the base frame at time $t$. To gain insight into the operation of the entropy model, visualizations of the corresponding latent features are also shown. As seen in these sample visualizations, 
incorporating the attention layer into the proposed entropy parameter estimation network facilitates its ability to selectively focus on important features within the input data, resulting in more efficient parameter estimation. For instance, the attention mechanism weakens static areas that correspond to the floor in the first channel of $\hat{\sigma}_t^e$; since these areas are static, $\hat{\sigma}_t^e$ reduces towards zero, while it remains higher near moving objects, the basketball and players. We can also observe that some parts of the last channel of $\hat{\mu}_t^e$ are mostly obtained from the last channel of the base features $\hat{y}_t^b$ (e.g., those in the solid circle) while some other parts (e.g., those in the dashed circle) are mostly taken from the enhancement features of the previous frame, $\hat{y}_{t-1}^e$. By utilizing attention and feeding it both the base and previous enhancement features, the entropy model is able to leverage the information available in the current base frame as well as previously-reconstructed frames in the enhancement layer, thereby enhancing its ability to achieve better compression of the current enhancement frame. \\

\textbf{Rate Estimation:}
As will be discussed in the following section, the base and enhancement layers are trained on sequences of contiguous frames to allow the models to learn temporal relationships. Specifically, the gradients of the current frame are back-propagated to the reference frames that preceded it within a temporal window of length $N$ frames. As a result, the tensors $\hat{y}_{t-1}^e$, $\hat{y}_t^b$ and $\hat{c}_t^e$ are endowed with substantial information not only from the current frame, but also from the preceding frames. This characteristic enables the system to take advantage of the correlations among the latent representations of multiple frames to model the entropy more effectively.

Having obtained $\hat{\mu}_t^e$, $\hat{\sigma}_t^e$, and $\theta_t^e=\{\hat{h}_t^e, \hat{c}_t^e, \hat{y}_{t-1}^e, \hat{y}_t^b\}$, the formulation for the conditional distribution of the latent representations 
can be obtained as follows \cite{balle2018}:
\begin{equation}
    \label{eq:prob_latent_enh}
\begin{split}
    p_{\hat{y}_t^e|\theta_t^e}(\hat{y}_t^e|\theta_t^e) & =  \prod_{i} p_{\hat{y}_t^e|\theta_t^e}(\hat{y}_{t,i}^e|\theta_t^e),
\end{split}  
\end{equation}
where the index $i$ represents the spatial location, and
\begin{equation}
    \label{eq:prob_latent_enh2}
\begin{split}
    p_{\hat{y}_t^e|\theta_t^e}(\hat{y}_{t,i}^e|\theta_t^e) & =  \left[(\mathcal{L}(\hat{\mu}_{t,i}^e, \hat{\sigma}_{t,i}^{e})*\mathcal{U}\left(-\frac{1}{2},\frac{1}{2}\right)\right](\hat{y}_{t,i}^e) \\
    & = \textrm{cdf}\left(\hat{y}_{t,i}^e+\frac{1}{2}\right) - \textrm{cdf}\left(\hat{y}_{t,i}^e-\frac{1}{2}\right),
\end{split}  
\end{equation}
where $\mathcal{L}(\hat{\mu}_{t,i}^e, \hat{\sigma}_{t,i}^{e})$ represents the Laplacian probability density function with mean $\hat{\mu}_{t,i}^e$ and variance $\hat{\sigma}_{t,i}^e$, $*$ represents the convolution, $\mathcal{U}(-\frac{1}{2},\frac{1}{2})$ represents the uniform probability density in the range $[-\frac{1}{2},\frac{1}{2}]$ (which models uniform quantization), and $\textrm{cdf}(\cdot)$ is the cumulative distribution function (CDF) of the resulting distribution. The estimated per-sample rate for coding $\hat{y}_t^e$ is given by $-\log_2(p_{\hat{y}_t^e|\theta_t^e}(\hat{y}_{t}^e|\theta_t^e))$, and the estimated entropy is the expectation of this quantity. This rate estimate is one of the loss terms used for training the enhancement layer, as discussed next.

\subsubsection{Enhancement Layer Loss Function}
The loss function for the enhancement layer is defined as:
\begin{equation}
    \label{eq:loss_enh}
    L_{enh}(t) = R_{enh}(t) + \lambda_{enh}~\textrm{MSE}(x_t, \hat{x}_t),
\end{equation}
where $x_t$ and $\hat{x}_t$ are the input and enhancement layer's decoded frame in RGB domain, respectively, at time $t$, $\lambda_{enh}$ is the Lagrange multiplier, and $R_{enh}(t)$ is the rate estimate for coding the current enhancement-layer frame, which is defined as follows:
\begin{equation}
    R_{enh}(t) = R_{enh}^{\textrm{motion}}(t) + R_{enh}^{\textrm{context}}(t),
\end{equation}
where $R_{enh}^{\textrm{motion}}(t)$ is the total rate estimate for coding the motion information in the enhancement layer and $R_{enh}^{\textrm{context}}(t)=\mathbb{E}[-\log_2(p_{\hat{y}_t^e|\theta_t^e}(\hat{y}_{t}^e|\theta_t^e))]$.




\section{Experiments}
\label{sec:experiments}
In this section, we describe the experiments performed to evaluate the performance of the proposed scalable video codec. Training details are described in Section~\ref{sec:training}, followed by the evaluation methodology in Section~\ref{sec:eval}, and the experimental results for the base and enhancement layers in Sections~\ref{sec:eval_base} and~\ref{sec:eval_enh}, respectively. Break-even analysis is presented in Section~\ref{sec:eval_bep}.

\subsection{Training}
\label{sec:training}
For training 
the proposed system, we used the VIMEO-90K Setuplet dataset~\cite{vimeo}, which consists of 91,701 7-frame sequences with fixed resolution $448\times 256$, extracted from 39K selected video clips. We randomly cropped
these clips into $256\times 256$ patches and used them to train the system. We used Adam~\cite{adam} optimizer with batch size of 4. 
We employed
a training strategy commonly used in recent papers~\cite{error1, canf, dcvc_tcm}, where the video codec is trained on a group of consecutive frames. In general, utilizing a larger group of frames in training can lead to better results. However, similar to~\cite{canf, dcvc_tcm} and motivated by memory constraints, in our experiments the loss function for each layer is computed over $N=5$ consecutive frames as follows:
\begin{equation}
\begin{split}
    L_{base}^* = \frac{1}{N} \sum_{t=1}^N L_{base}(t), \quad \,
    L_{enh}^* = \frac{1}{N} \sum_{t=1}^N L_{enh}(t),
\end{split}
\end{equation}
where $L_{base}(t)$ and $L_{enh}(t)$ are defined as in (\ref{eq:base_loss}) and (\ref{eq:loss_enh}), respectively. Using this training strategy, the gradients of each frame are back-propagated towards other frames in the group, which allows the system to capture temporal relationships among a group of consecutive frames.

For the base layer, we used $\lambda_{base} \in \{2,4,8,16\}$. All networks in the base layer were first initialized with their pre-trained weights from~\cite{canf}. For YOLOv5, we used the PyTorch implementation provided in~\cite{ultra}, and used the checkpoint for the ``small'' model for the initialization. The learning rate was set to a small value $10^{-6}$, and the entire layer was trained end-to-end for 10 epochs. We first trained the base layer, and then kept it frozen during the training of the enhancement layer.

For the enhancement layer, we used $\lambda_{enh} \in \{256,512,1024,2048\}$. The learning rate was set to $10^{-4}$ for the first 5 epochs, and then reduced to $10^{-5}$ for the next 10 epochs. We first trained the enhancement layer with $\lambda_{enh}=2048$ (highest rate), then all lower-rate
models were initialized from this model.

\subsection{Evaluation Methodology}
\label{sec:eval}
\subsubsection{Evaluation Datasets}
To evaluate a video codec intended for both object detection and input frame recovery, it is necessary to test it on datasets consisting of raw (uncompressed) video and associated object labels. To our knowledge, only two such datasets are currently available -- SFU-HW-Objects-v1~\cite{sfu_hw} and TVD~\cite{tvd} -- and both are being used in the MPEG-VCM standardization effort~\cite{mpeg_vcm}. The SFU-HW-Objects-v1 dataset is based on raw YUV420 video sequences that were used in the development of video coding standards such as HEVC~\cite{hevc} and VVC~\cite{vvc}. These sequences are usually referred to as ``HEVC test sequences'' and come in several classes. We used sequences from three classes: Class B (1920$\times$1080 resolution), Class C (832$\times$480), and Class D (416$\times$240). Each sequence contains between 240--600 frames and their frame rates range from 24--50 frames per second. A subset of these sequences was labeled with COCO7-style object labels. The TVD dataset includes 86 sequences, each consisting of 65 frames. However, object labels are only provided for 55 unique video triplets (3-frame sequences). 

\begin{figure*}
\centering
\includegraphics[width=\textwidth]{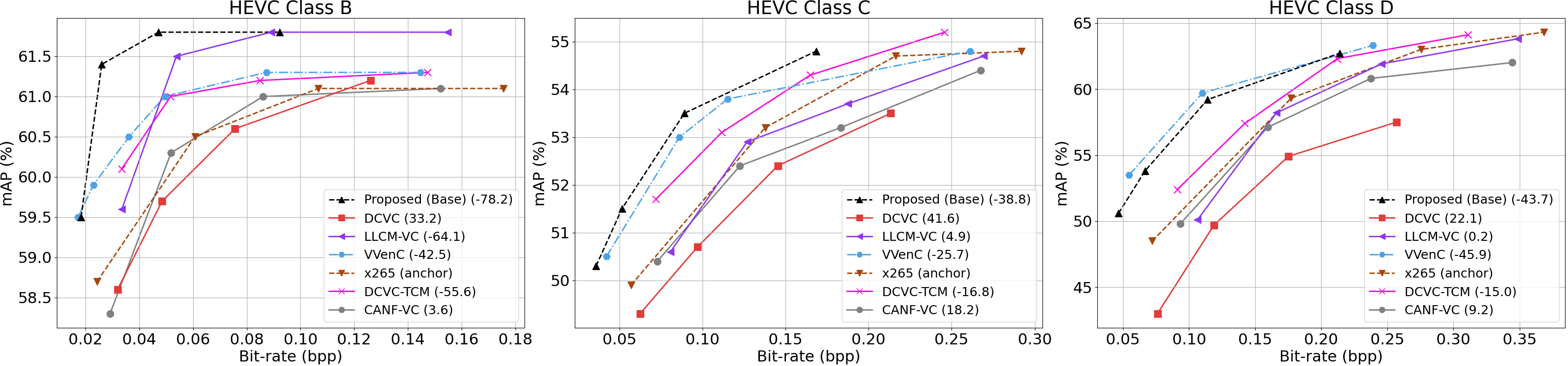}
\caption{Comparing the performance of various methods in terms of object detection mAP (\%) on three datasets. The straight horizontal line in each graph is the performance of YOLOv5 without input compression. In the legend of each graph, the BD-Rate(\%) of the corresponding method is also shown with respect to VTM (anchor).}
\label{fig:results_map}
\end{figure*}

\begin{figure*}[t]
\centering
\includegraphics[width=\textwidth]{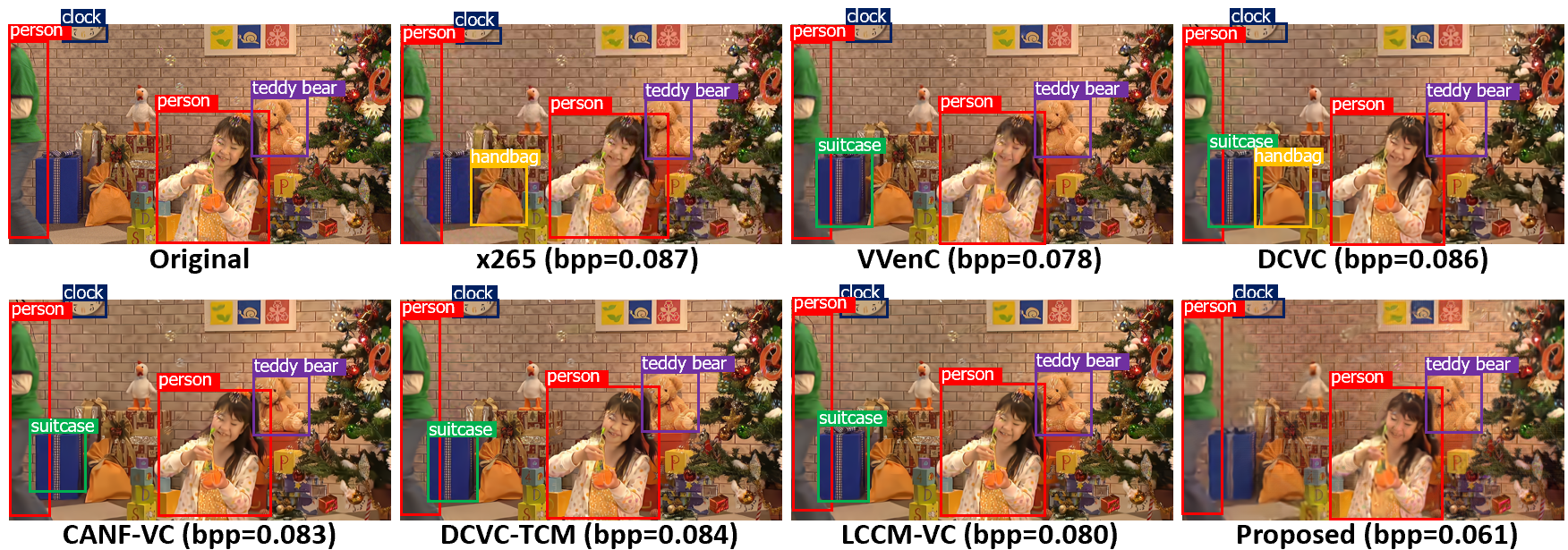}
\caption{Visualizing object detection performance of various methods on a sample frame from PartyScene (HEVC Class C). The ground-truth object labels come from the SFU-HW-Objects-v1 dataset.}
\label{fig:base_visual}
\end{figure*}

\subsubsection{Benchmark Methods}
We used the following video codecs as benchmarks: HEVC, VVC, DCVC-TCM~\cite{dcvc_tcm}, CANF-VC~\cite{canf}, and DCVC \cite{dcvc}. For HEVC, we used HM 18.0~\cite{hm} with LDP profile (encoder\_lowdelay\_P\_main.cfg), and for VVC, we employed VTM 23.4 \cite{vtm} with LDP profile (encoder\_lowdelay\_P\_vtm.cfg). Among these, VTM was chosen as the anchor against which all other codecs are compared. For HM, which is the standard implementation of HEVC~\cite{hevc}, the quantization parameter (QP) values were $\{27, 29, 32, 38\}$. For VTM, which is the standard implementation of VVC~\cite{vvc}, QP values were $\{22, 24, 27, 29\}$. The internal color space for HM and VTM was YUV420 (8 bit). Also, four past reference frames were used for motion estimation in these conventional codecs. However, all the tested learned codecs (CANF-VC, DCVC-TCM, LCCM-VC, DCVC, and the proposed method) use only one reference frame in their optical flow estimators.  As reported in \cite{dcvc_tcm}, if we use only one reference frame for the conventional codecs instead of four, their performance will be degraded considerably. In line with the common practice in previous learned video codecs \cite{canf,dcvc_tcm}, in the experiments, we set the IntraPeriod (also known as Group Of Pictures, GOP) to 32 for all codecs when testing them on SFU-HW-Objects-v1 to be able to compare our results with other existing works. For intra-frame coding of the first frame in each GOP, we used the intra-frame coder from CANF-VC for the base layer and the intra-frame coder from DCVC-TCM for the enhancement layer. On the TVD dataset, the GOP size is set to 3 because sequences are 3-frames long. For CANF-VC, DCVC-TCM, LCCM-VC, and DCVC, which require RGB frames, we first converted all YUV 420 videos to RGB format using ffmpeg \cite{ffmpeg}, following the ITU-R BT.709 recommendation \cite{ITU709}. The resulting RGB frames were then used for both training and inference.

\subsubsection{Evaluation Metrics}
Following the existing practice in the learned video coding literature~\cite{canf, mlvc, dcvc, dcvc_tcm}, to evaluate the performance of various codecs, the bitrates were measured in bits per pixel (bpp) and the frame reconstruction quality was measured by RGB Peak Signal-to-Noise Ratio (PSNR-RGB) and RGB Multi-Scale Structural Similarity Index Metric (MS-SSIM-RGB)~\cite{ssim}. Object detection accuracy was measured by Mean Average Precision (mAP)~\cite{yolo3}. The mAP is a popular metric used to evaluate the accuracy of object detection models. It is computed by first calculating the Average Precision (AP) for each class, which is the area under the Precision-Recall curve. Precision is the ratio of correctly predicted positive samples to all predicted positives, while Recall is the ratio of correctly predicted positives to all actual positives. The mAP is then obtained by averaging the APs across all classes, giving an overall performance score for the model's ability to detect objects across different categories. Following the methodology used in MPEG-VCM~\cite{mpeg_vcm}, these quantities can be agglomerated into a single numerical score called Bj\o{}ntegaard Delta Rate (BD-Rate)~\cite{Bjontegaard}, which computes the average bitrate difference between two codecs at the equivalent quality level, where the quality may be measured by PSNR-RGB, MS-SSIM-RGB, or mAP, as the case may be. So, for example, if a certain codec achieves --10\% BD-Rate-mAP relative to a given anchor, it means that it achieves, on average, 10\% bitrate reduction relative to the anchor while producing the same mAP.

\subsection{Object Detection Performance}
\label{sec:eval_base}
First, we evaluate the object detection performance. For the proposed codec, we only need to decode the base-layer bitstream and feed it to the pre-trained YOLOv5 back-end. For other codecs, we decode each frame and then feed it to the pre-trained YOLOv5. 
Fig.~\ref{fig:results_map} shows the mAP (\%) vs. bitrate (bpp) on the three classes of sequences from the SFU-HW-Objects-v1 dataset. The straight horizontal line in each graph is the performance of YOLOv5 without input frame compression.
The legend in each graph shows BD-Rate-mAP percentage in the brackets, relative to the VTM anchor. Negative values indicate bit savings relative to the anchor at the equivalent mAP. BD-Rate-mAP results are also summarized in Table~\ref{tab:base_bd}, where the best result in each row is highlighted in bold. The table also includes the results on the TVD dataset. However, since the sequences in the TVD datset are short (3 frames each), the performance of most codecs is similar. The TVD sequences seem too short to allow distinguishing the various codecs' ability to capture long-range relationships among features relevant to object detection. The last two rows in Table~\ref{tab:base_bd} show the average BD-Rate-mAP results, averaged according to the number of frames in each dataset. We see that the overall performance (bottom row) is dominated by the performance on SFU-HW-Objects-v1, since this dataset has many more frames than TVD.  
Overall, the results indicate that the proposed codec in the base layer outperforms all other codecs. 
HM comes as the second best, followed by LCCM-VC and DCVC-TCM. 

Fig.~\ref{fig:base_visual} provides a visual example of object detection on a sample frame from the PartyScene sequence (HEVC Class C). In this figure, we show the bounding box of the detected objects overlaid on the reconstructed frame. For the proposed method, we show the base frame because only base-layer information is used. As seen in this example, the proposed method is capable of achieving better detection accuracy at a lower rate compared to other methods. Specifically, we note that in the frames produced by other codecs, the gift box is often mis-detected as a suitcase, and the gift bag is confused with a handbag. While these objects might be similar to some extent, the compression artifacts introduced by other codecs cause the detector to produce incorrect detections, whereas our method preserves the relevant features well, avoiding mis-detection even at lower rates than other methods.

  \begin{sidewaystable*}
     \caption{Object detection performance in terms of BD-Rate-mAP (\%) on different datasets. The anchor is VTM. ``Base" represents the proposed method in the base layer.}\label{tab:base_bd}
     \centering
    \begin{tabular}{c|c|c|c|c|c|c}
        \toprule
       Dataset  & CANF-VC \cite{canf} & DCVC-TCM \cite{dcvc_tcm}& LCCM-VC \cite{lccm} & DCVC \cite{dcvc} & Base & HM\\
       \midrule
         SFU-HW-Objects-v1 (HEVC Class B) &241.5 & 86.5&10.6&325&\textbf{--34.2}&9.5\\ \addlinespace[1pt] \hline \addlinespace[1pt]
         SFU-HW-Objects-v1 (HEVC Class C)& 54.8 & 8.3& 37.1& 84.9& \textbf{--19.8}& 38.7\\ \addlinespace[1pt]
         \hline \addlinespace[1pt]
         SFU-HW-Objects-v1 (HEVC Class D)&  102.3&48.3&76.7&175&\textbf{4.4}&78.1 \\ \addlinespace[1pt]
         \hline \addlinespace[1pt]
         TVD & 0.1& 0.2&0.1&33.1&\textbf{-0.5}&0.2\\
         \midrule[0.6pt]
         Average (SFU-HW-Objects-v1) & 136.2&48.8&40.5&198.9&\textbf{-17.1}&41.1\\
         \midrule[0.6pt]
         Average & 134.2&48.18&39.9&196.5&\textbf{-16.8}&43.5 \\
         \bottomrule
    \end{tabular}

    \vspace{0.5\baselineskip}
    \caption{Frame reconstruction performance in terms of BD-Rate-PSNR (\%) on different datasets. The anchor is VTM. The last three columns from the right show the results of the three cases of the proposed method. }\label{tab:enh_psnr}
     \centering
    \begin{tabular}{c|c|c|c|c|c|c|c|c}
        \toprule
       Dataset  & CANF-VC \cite{canf} & DCVC-TCM \cite{dcvc_tcm} & LCCM-VC \cite{lccm} & DCVC \cite{dcvc}& Base & Enh& Base+Enh & HM\\
       \midrule
         HEVC Class B& 59.8&24.1&55.5&101.7&162.4&29.1&65.3&\textbf{23.4}\\
         \addlinespace[1pt] \hline \addlinespace[1pt]
         HEVC Class C& 56.1&36.9&52.0&99.4&136.5&32.5&68.5&\textbf{11.6}\\
         \addlinespace[1pt] \hline \addlinespace[1pt]
         HEVC Class D& 58.1&30.4&53.6&30.4&133.7&26.7&62.4&\textbf{17.4}\\
         \addlinespace[1pt] \hline \addlinespace[1pt]
         TVD&27.5&31.4&25.6&67.1&83.4&19.6&44.5&\textbf{9.6}\\
         \midrule[0.6pt]
        Average (HEVC)&  58.0&30.3&53.7&77.9&144.7&29.4&65.4&\textbf{17.6}\\

         \midrule[0.6pt]
         Average & 57.6&30.3&53.3&77.7&143.9&29.3&65.1&\textbf{17.5}\\
         \bottomrule
    \end{tabular}

    \vspace{0.5\baselineskip}
        \caption{Frame reconstruction performance in terms of BD-Rate-MS-SSIM (\%) on different datasets. The anchor is VTM. The last three columns from the right show the results of the three cases of the proposed method.} \label{tab:enh_ssim}
    \begin{tabular}{c|c|c|c|c|c|c|c|c}
        \toprule
       Dataset  & CANF-VC \cite{canf} & DCVC-TCM \cite{dcvc_tcm} & LCCM-VC \cite{lccm} & DCVC \cite{dcvc}&Base&Enh &Base+Enh&HM\\
        \addlinespace[1pt] \hline \addlinespace[1pt]
         HEVC Class B  & 56.7&1.2&39.1&72.3&60.8&\textbf{3.6}&34.4&9.2\\
        \addlinespace[1pt] \hline \addlinespace[1pt]
         HEVC Class C  &51.8&6.7&44.1&59.5&51.1&\textbf{4.6}&35.5&37.9\\
        \addlinespace[1pt] \hline \addlinespace[1pt]
         HEVC Class D  & 70.1&11.2&61.7&56.3&53.4&\textbf{0.3}&33.5&41.0\\
        \addlinespace[1pt] \hline \addlinespace[1pt]
         TVD  & 15.1&8.2&13.3&9.6&39.3&\textbf{0.7}&16.2&14.6\\
         \midrule[0.6pt]
         Average (HEVC)&  59.4&6.2&48.0&63.0&55.3&\textbf{2.8}&34.4&28.7\\
        \midrule[0.6pt]
         Average&  58.8&6.2&47.5&62.2&55.0&\textbf{2.8}&34.1&28.5\\
         \bottomrule
    \end{tabular}
   \end{sidewaystable*}

\subsection{Frame Reconstruction Performance}
\label{sec:eval_enh}
When it comes to the performance of reconstructing compressed frames for human viewing, the situation is straightforward for conventional video codecs, whether learned or handcrafted: we simply decode the compressed bitstream, generate the decoded frames, and measure their quality using a metric (PSNR-RGB or MS-SSIM-RGB in this work). However, for the proposed codec, the following options can be considered:

\begin{figure*}
\centering
\includegraphics[width=\textwidth]{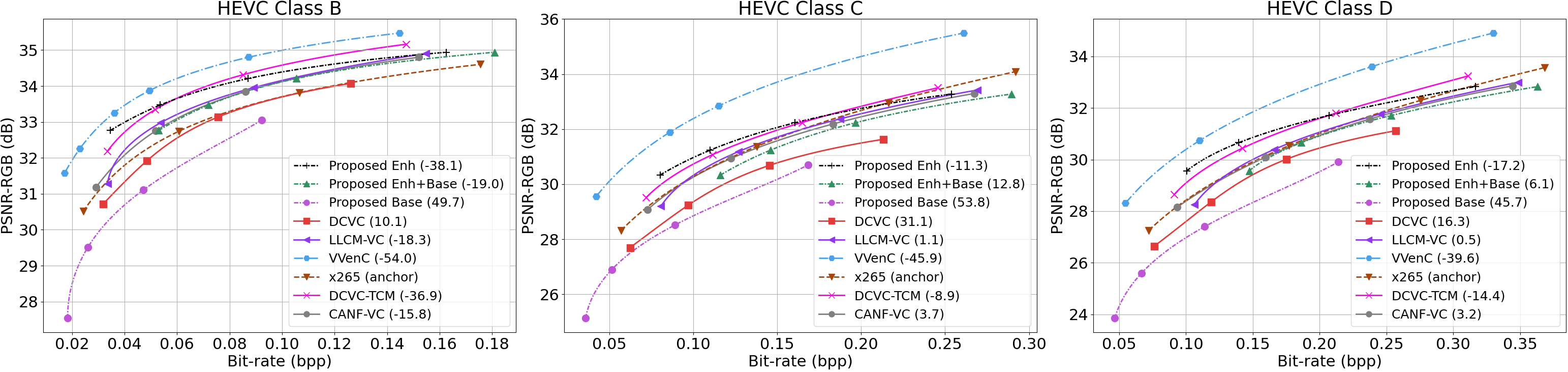}
\caption{Comparing the input reconstruction performance of various methods in terms of PSNR-RGB (dB) on three datasets. In the legend of each graph, the BD-Rate(\%) of the corresponding method is also shown with respect to VTM (anchor). ``Base," ``Enh," and ``Base+Enh" represent different cases of the proposed system.}
\label{fig:results_psnr}
\end{figure*}

\begin{figure*}
\centering
\includegraphics[width=\textwidth]{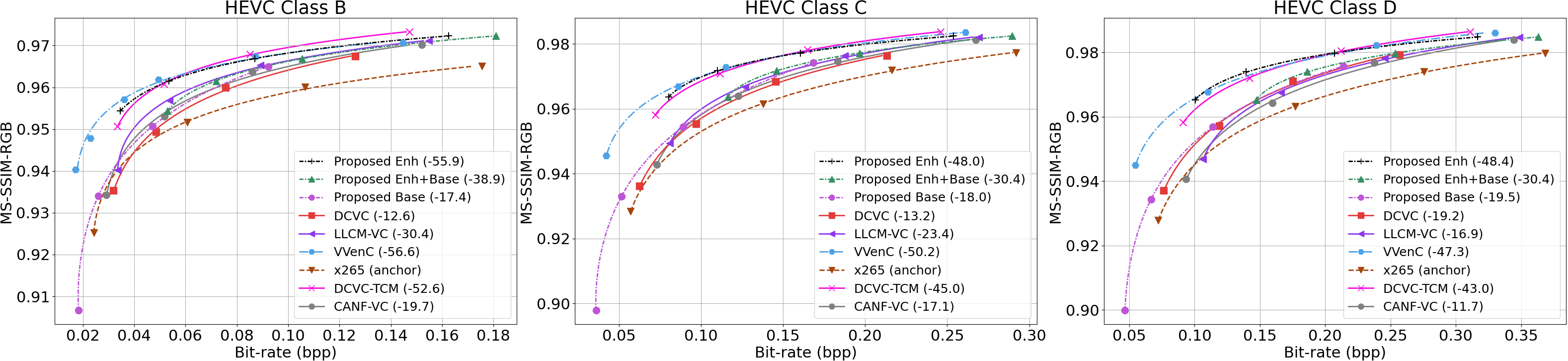}
\caption{Comparing the input reconstruction performance of various methods in terms of MS-SSIM-RGB on three datasets. In the legend of each graph, the BD-Rate(\%) of the corresponding method is also shown with respect to VTM (anchor). ``Base," ``Enh," and ``Base+Enh" represent different cases of the proposed system.}
\label{fig:results_ssim}
\end{figure*}

\begin{itemize}
    \item \textbf{Option 1:} First, note that the base layer of the proposed system generates video frames and stores them in the base frame buffer, as shown in Fig.~\ref{fig:flowchart_base_layer}. Such frames are intended for object detection rather than human viewing; they preserve enough information to detect objects at low rates, but lack some of the information that is irrelevant for object detection.
    Nevertheless, we could ask what their quality is in terms of frame reconstruction metrics. In Figs.~\ref{fig:results_psnr}-\ref{fig:results_ssim} and Tables~\ref{tab:enh_psnr}-\ref{tab:enh_ssim}, we refer to this approach as ``\textbf{Base}''. In this case, only the base-layer bitrate is counted, since base frames are generated from the base-layer bitstream only.

    \item \textbf{Option 2:} The intended use of the proposed codec is in a closed-loop automated analytics system, where analysis is performed using object detection (base layer) information, and frame reconstruction is initiated depending on the results of the base-layer analysis. For example, in a traffic monitoring system, object detection runs continuously. When the arrangement of detected objects suggests a traffic violation, the enhancement layer is turned on to provide additional information for frame reconstruction and human viewing. Hence, given that base-layer information is already available at the analysis server, only the enhancement layer bitstream needs to be communicated to provide high-quality frame reconstruction. The bitrate in this case is the enhancement-layer bitrate, and the frame quality is measured on the enhancement-layer frames. This approach is termed ``\textbf{Enh}'' in Figs.~\ref{fig:results_psnr}-\ref{fig:results_ssim} and  Tables~\ref{tab:enh_psnr}-\ref{tab:enh_ssim}. It is worth noting that the MPEG-VCM evaluation framework~\cite{mpeg_vcm_evaluation} suggests comparing this approach against conventional coding approaches. The goal of MPEG-VCM is stated as~\cite{mpeg_vcm_evaluation}: ``\textit{The bitrate of the additional compressed bitstream shall be less than the bitrate of the bitstream at similar quality as measured by PSNR, which is the output of the VVC encoding of the unprocessed video.}''   

    \item \textbf{Option 3:} Finally, we can also count the total bitrate (base + enhancement layer) for the proposed codec, along with enhancement-layer frame quality. This is referred to as  ``\textbf{Base+Enh}'' in Figs.~\ref{fig:results_psnr}-\ref{fig:results_ssim} and  Tables~\ref{tab:enh_psnr}-\ref{tab:enh_ssim}. It is debatable whether the comparison of conventional coding against this case is realistic. Such a comparison may make sense if we knew \emph{in advance} that a situation of interest (e.g., traffic violation in our earlier example) would occur at some point in time, without performing any visual analysis. In such a case, our codec would still have to encode both base and enhancement layer bitstreams, even though the result of the analysis is apparently known ahead of time. Nevertheless, we included these ``oracle-style'' results for completeness, even though they show less favorable performance of our codec. 
\end{itemize}

\begin{figure*}[t]
\centering
\includegraphics[width=\textwidth]{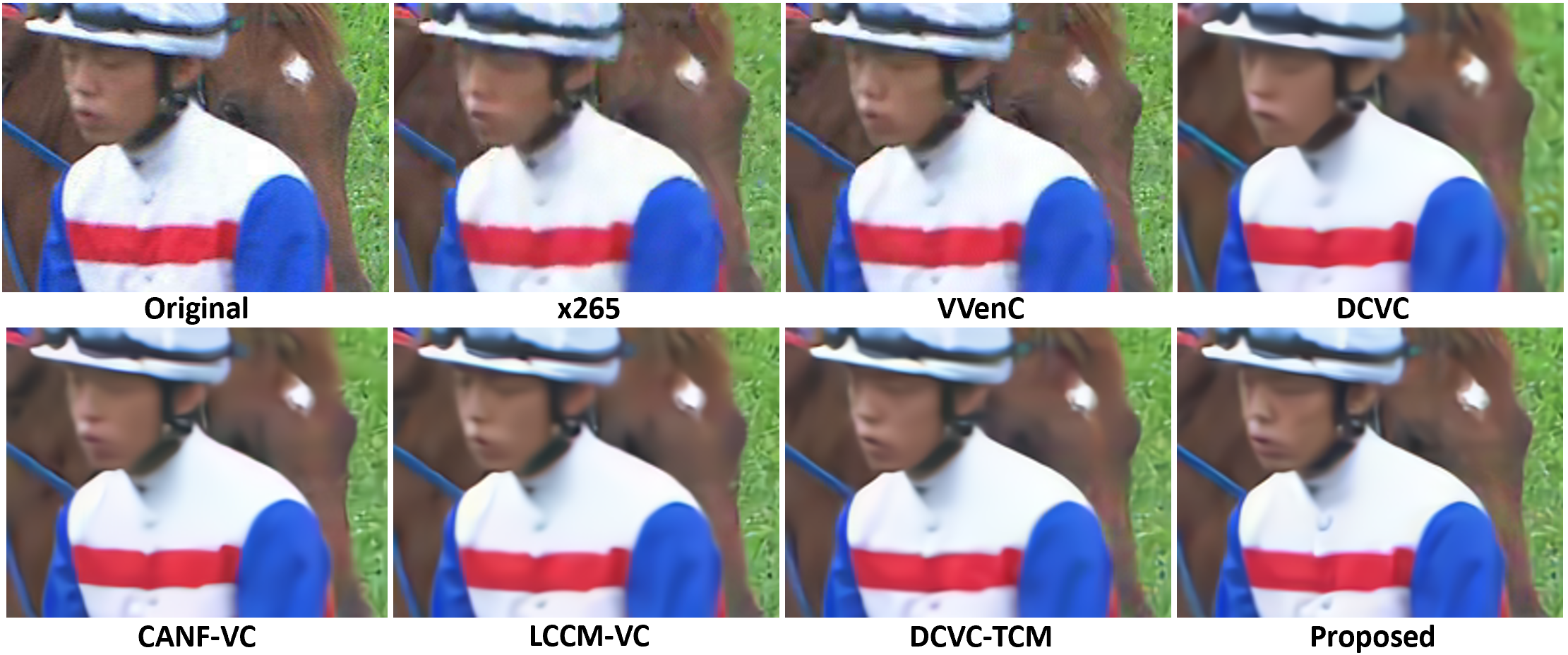}
\caption{Comparing the input reconstruction performance of various methods on a patch from a sample frame of RaceHorses (HEVC Class C). The (bpp, PSNR-RGB, MS-SSIM-RGB) for HM, VTM, DCVC, CANF-VC, DCVC-TCM, LCCM-VC, the proposed method (Base+Enh), and the proposed method (Enh) is (0.43,33.64,0.9759), (0.43,33.96,0.9807), (0.44,32.65,0.9678), (0.42,33.50,0.9750), (0.43,33.61,0.9760), (0.44,33.78,0.9771), (0.42,33.60,0.9758), and (0.37,33.60,0.9758), respectively. Note that the proposed Enh and the proposed Base+Enh produce the same frame -- indicated by ``Proposed'' in the figure -- only the bits are counted differently.}
\label{fig:enh_visual}
\end{figure*}

We now evaluate the proposed method using these three options and compare its performance with other benchmark methods through both quantitative and qualitative analyses.

\subsubsection{Quantitative Results using PSNR-RGB}
Fig.~\ref{fig:results_psnr} shows the PSNR-RGB (in dB) versus bitrate for various cases on the three classes of HEVC video sequences. The legend shows BD-Rate-PSNR (\%) relative to VTM in the brackets. 
These results are also tabulated in Table~\ref{tab:enh_psnr}, along with BD-Rate-PSNR on the TVD dataset. As seen from these results, for the input frame reconstruction task, the best-performing method is VTM (as all values in this table are positive, meaning that all methods performs worse than VTM) followed by HM and then our proposed Enh. The gap between VTM and our Enh is about 29\%. This is not surprising, given that conventional video codecs are known to perform very well on the PSNR metric. It is interesting to note that even though our Base method uses the base layer bitstream only (seemingly a low bitrate), it performs the worst among all the methods tested in Table~\ref{tab:enh_psnr}. This is because BD-Rate-PSNR between a given codec and the anchor is measured at the equivalent PSNR, and our Base method needs a lot of bits to reach the equivalent PSNR to VTM. Our base layer was designed to support object detection, so it lacks the details needed to achieve high PSNR, and therefore needs to use a lot of bits to bring those details back in.

\subsubsection{Quantitative Results using MS-SSIM-RGB} 
Fig.~\ref{fig:results_ssim} and Table~\ref{tab:enh_ssim} show analogous results but with MS-SSIM-RGB as the frame quality metric. 
Learned codecs are known to perform better on MS-SSIM-RGB than on PSNR-RGB, even when MSE is used in their loss function~(\ref{eq:loss_enh}). Indeed, this is seen in Fig.~\ref{fig:results_ssim} and Table~\ref{tab:enh_ssim}. 
Specifically, although VTM still comes out as the best codec on BD-Rate-SSIM, the gap between it and our Enh (the second-best method) is now about 2.8\%. DCVC-TCM (the third-best method) also comes to within a few percentage points of VTM. Almost all other codecs perform comparatively better in Table~\ref{tab:enh_ssim} than in Table~\ref{tab:enh_psnr}. It is interesting to note that our Base method records the highest jump in performance between Table~\ref{tab:enh_psnr} and Table~\ref{tab:enh_ssim}. This suggests that object-relevant information, which is preserved in our base layer, is also favorable to quality metrics such as MS-SSIM-RGB.

\subsubsection{Qualitative Results}
\begin{figure*}[t]
\centering
\includegraphics[width=\textwidth]{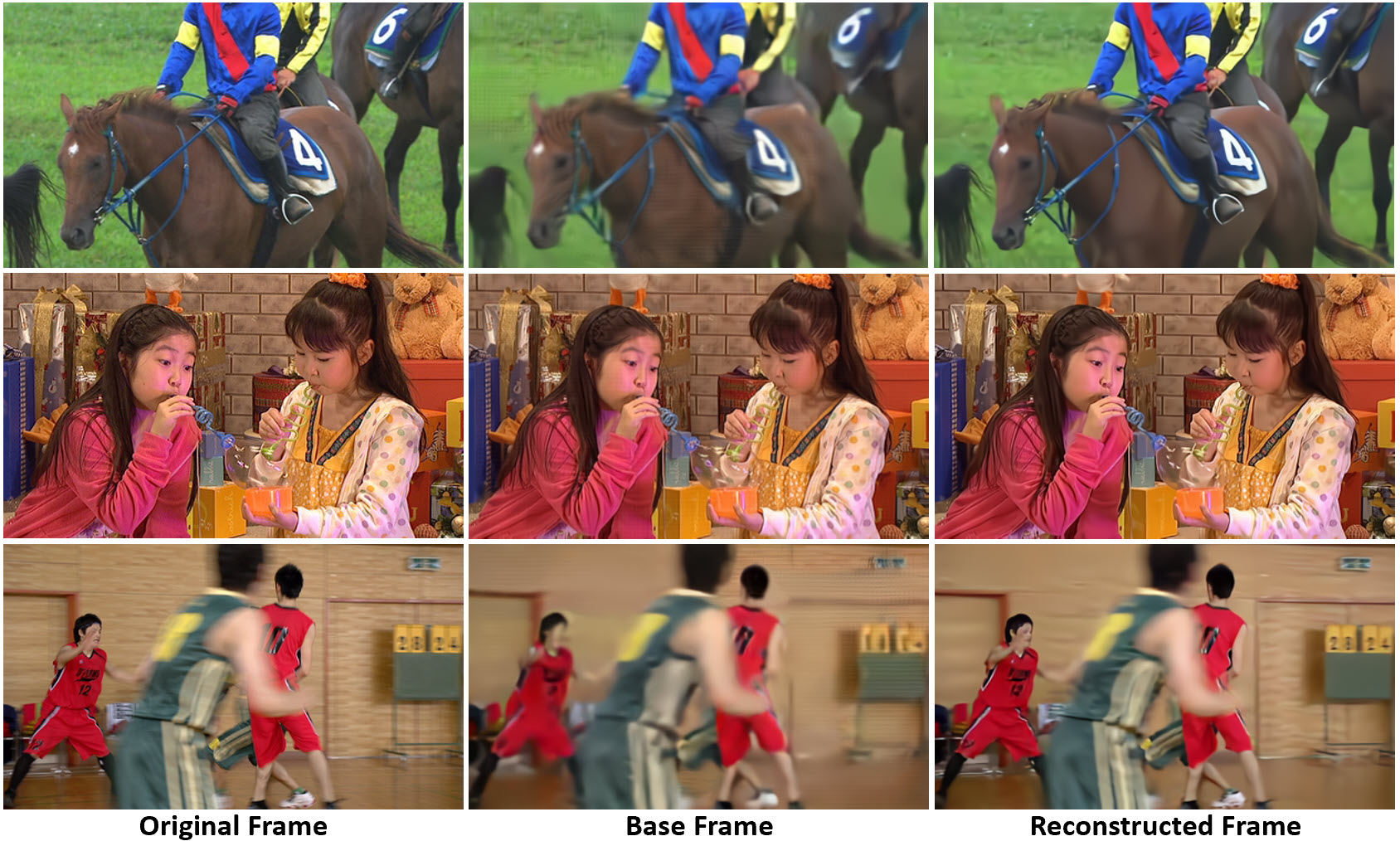}
\caption{Three examples of the visual quality of the base frame and the corresponding enhancement-layer reconstructed frame. Top row: a frame from RaceHorses (HEVC Class D, Base Rate=0.13, Enhancement Rate=0.05); Middle row: a frame from BasketballPass (HEVC Class D, Base Rate=0.07, Enhancement Rate=0.06); Bottom row: a frame from the TVD dataset (Base Rate=0.16, Enhancement Rate=0.08). }
\label{fig:visuals}
\end{figure*}

Fig.~\ref{fig:enh_visual} shows a patch from a sample frame from RaceHorses (HEVC Class C) reconstructed by various codecs at a similar bitrate: 0.42--0.44 bpp for all codecs, except our Enh at 0.37 bpp. As seen from these examples, learned codecs tend to suppress some details, especially in the grass in the background; among the learned codecs, LCCM-VC and the proposed one seem to have recovered most of these details.  Fig.~\ref{fig:visuals} shows visual comparisons of base and enhancement-layer (reconstructed) frames produced by our codec. It is evident that base frames lack detail, but they preserve enough detail needed to identify objects, which is what the base layer is designed for.

\subsection{Break-Even Analysis}
\label{sec:eval_bep}
In the preceding sections, we provided experimental results for the machine task (object detection) and human viewing separately. We saw that the proposed scalable codec performs better than the alternatives on the machine task (Table~\ref{tab:base_bd}) while VTM is the best-performing codec on frame reconstruction for human viewing (Tables~\ref{tab:enh_psnr} and~\ref{tab:enh_ssim}). 

So, which codec is better -- VTM or the proposed one? That depends on how frequently one expects human viewing will be required in a particular application: whenever human viewing is needed, VTM is better, but when the system performs the machine task only, the proposed codec is better. To enable numerical comparison of codecs in such a scenario, we adopt \emph{break-even analysis} from~\cite{hyomin_svc}. The gist of break-even analysis is as follows. From Table~\ref{tab:base_bd}, the proposed codec is, on average, 16.8\% more efficient than VTM when the machine task is used. In other words, it uses 0.832 (or 83.2\%) of the bits used by VTM in this case. Meanwhile, from Table~\ref{tab:enh_psnr}, the proposed codec in the Enh regime is, on average, 29.3\% less efficient than VTM (it uses 1.293 of the VTM bits) on frame reconstruction for human viewing, when frame quality is measured by PSNR-RGB.  
Let $\phi \in [0, 1]$ be the fraction of time that input frame reconstruction is needed for human viewing. The
amount of bits used by our system will be less than or equal
to that used by VTM if

\begin{equation}
\label{eq:bep}
    (1-\phi)\cdot0.832 + \phi \cdot 1.293 \leq 1.
\end{equation}
\color{black}
Solving for $\phi$ that achieves equality in (\ref{eq:bep}), we obtain the break-even point of $\phi = 0.37$. That is to say, if input reconstruction for human viewing is needed less than 37\% of the time, our system will provide overall bit savings over  VTM. However, if MS-SSIM-RGB is used as the frame quality metric, we obtain  $\phi=0.86$. Using this analysis, the proposed method in the Enh regime always outperforms ($\phi=1$) the other codecs in this study, meaning that it provides bit savings regardless of the fraction of time used for human viewing. For completeness, Table~\ref{tab:bep} also shows the break-even points of Base+Enh against VTM. Although Base+Enh results are less favorable to our method and less realistic in a practical setting, as discussed earlier, even in this case our method provides gains over VTM when human viewing is needed less than 21\% of the time. 

But how often is human viewing needed in these kinds of applications - traffic monitoring, surveillance, etc.? We do not have a precise answer to this question. One reference point could be a report published in 2015~\cite{IPVM2015}, suggesting that less than 1\% of all surveillance video is watched live. Given the proliferation of surveillance technology and likely much slower increase (if any) of human viewers of surveillance video, the percentage is almost certainly lower now. Although ``human viewing'' in our system is not the same as ``live viewing,'' the figure of 1\% ($\phi=0.01$) can serve as a ballpark estimate, and we see in Table~\ref{tab:bep} that the break-even points of our system are many-fold higher than that. In other words, our system could comfortably provide gains over VTM in a practical surveillance scenario. 

\begin{table}[t]
    \centering
    \caption{The break-even point $\phi$ against VTM.}
    \begin{tabular}{c|c|c}
        \toprule
       Frame quality metric  & Enh & Base+Enh\\
        \addlinespace[1pt] \hline \addlinespace[1pt]
         PSNR-RGB  & 0.37 & 0.21\\
        \addlinespace[1pt] \hline \addlinespace[1pt]
         MS-SSIM-RGB & 0.86 & 0.33\\
         \bottomrule
        
    \end{tabular}
    \label{tab:bep}
\end{table}

\subsection{Ablation Study}
\label{sec:ablation}
In this section, we provide the results of an ablation study to demonstrate the efficiency of our proposed system. In our ablation study, we considered the following cases:
\begin{itemize}
    \item \textbf{Case 1}: Reusing the motion flow map of the base layer for the enhancement layer: In our proposed system, motion estimation is performed independently for both the base and enhancement layers, with the resulting motion flow maps encoded and transmitted separately to the decoder. One potential optimization is to share the base layer’s motion flow map with the enhancement layer, eliminating the need for a separate motion bitstream in the enhancement layer. We tested this approach on HEVC Class B. However, the results showed that using the base layer's flow map for the enhancement layer degraded the BD-Rate-PSNR performance by approximately 1\%. Therefore, we opted to maintain separate motion estimation and coding networks for each layer.
    \item \textbf{Case 2}: Freezing the cloned front-end network in the base layer: As discussed in Section \ref{sec:task}, during the training of the base layer, the cloned front-end network, $F^{\textrm{front-end}}_{\textrm{trainable}}(.)$, is allowed to be trained. A possible consideration is whether freezing this network might yield better results. We tested this approach on HEVC Class B, and the results showed that training this network led to approximately 2\% better rate savings. Therefore, in our proposed system, we did not freeze the weights of this network during training.
    \item \textbf{Case 3}: The role of the proposed entropy model: As discussed in Section \ref{sec:entropy_model}, our proposed entropy estimation module enhances the performance of the entropy model proposed in \cite{dcvc_tcm} by fusing the hyper priors of the current frame, $\hat{h}_t^e$, with the temporal prior $\hat{c}_t^e$, the decoded latents from the previous enhancement frame $\hat{y}_{t-1}^e$, and the base features of the current frame $\hat{y}_t^b$ through an entropy parameter estimation network. In contrast, the approach in \cite{dcvc_tcm} feeds only the hyper-prior $\hat{h}_t^e$ and the temporal prior $\hat{c}_t^e$ to the entropy model. Our design, however, enriches the input information by including $\hat{y}_t^b$ and $\hat{y}_{t-1}^e$ in the entropy model. Additionally, we incorporate an attention layer to enable the system to learn where to focus for effective entropy modeling. To evaluate the effectiveness of our proposed entropy estimation model, we replaced it with the original entropy model proposed in \cite{dcvc_tcm} and tested the performance of the enhancement layer on HEVC Class B. The results showed that our proposed entropy model yields approximately 1.5\% better rate savings in the enhancement layer.
\end{itemize}

\section{Conclusion} 
\label{sec:conclusions}
In this manuscript, we provided a theoretical justification for conditional lossy coding, a technique whose rate-distortion bounds are better than those of the commonly-used residual coding. We then presented the first end-to-end learned scalable video codec that employs conditional coding and is capable of efficiently supporting both human and machine vision. Specifically, object detection is supported by the codec's base layer, while input reconstruction for human viewing is supported by both the base and enhancement layer. 
We tested the proposed codec on the video object detection datasets used in MPEG-VCM, and compared it against recent learned and conventional video codecs. The results show that the proposed codec outperforms existing video codecs on the base task (object detection) while providing comparable performance to the best codecs on input frame reconstruction. We also performed break-even analysis against VTM, the best-performing codec on input frame reconstruction, which showed that in a practical scenario, our codec is able to provide bit savings against VTM if human viewing is needed less than 21\% of the time.  

\section*{Abbreviations}
\begin{itemize}
    \item DNN: Deep Neural Network
    \item HEVC: High Efficiency Video Coding
    \item VVC: Versatile Video Coding
    \item VCM: Video Coding for Machines
    \item CV: Computer Vision
    \item CANF-VC: Conditional Augmented Normalizing Flows for Video Coding
    \item LCCM-VC: Learned Conditional Coding Modes for Video Coding
    \item TCM: Temporal Context Mining
\end{itemize}


%





\ifCLASSOPTIONcaptionsoff
  \newpage
\fi



%



\bibliographystyle{IEEEtran}
\bibliography{refs}

%








\end{document}